\documentclass[11pt]{article}

\newcommand{\be}{\begin{equation}}
\newcommand{\ee}{\end{equation}}
\newcommand{\ba}{\begin{eqnarray}}
\newcommand{\ea}{\end{eqnarray}}
\newcommand{\nn}{\nonumber}
\newcommand{\ra}{\rightarrow}
\newcommand{\dd}{\mathrm d}

\newcommand{\ns}{\normalsize}

\begin{document}

\title{Coherent Radio Pulses From GEANT Generated Electromagnetic
Showers In Ice}

\author{{\bf Soebur Razzaque,$^a$ Surujhdeo Seunarine,$^b$}
\\
{\bf David Z. Besson,$^a$ Douglas W. McKay,$^a$ John P. Ralston,$\;^a$}
\\
{\bf and David Seckel$\;^c$}
\\ \\
{\it \ns $^a$Department of Physics \& Astronomy} \\
{\it \ns University of Kansas, Lawrence, KS 66045, USA} \\
{\it \ns $^b$ Department of Physics} \\
{\it \ns University of Canterbury, Private Bag 4800, Christchurch,
New Zealand} \\ 
{\it \ns $^c$ Bartol Research Institute} \\
{\it \ns University of Delaware, Newark, DE 19716, USA}}
\date{}
\maketitle

\begin{abstract}
Radio Cherenkov radiation is arguably the most efficient mechanism for
detecting showers from ultra-high energy particles of 1 PeV and above.
Showers occuring in Antarctic ice should be detectable at distances up
to 1 km.  We report on electromagnetic shower development in ice using
a GEANT Monte Carlo simulation.  We have studied energy deposition by
shower particles and determined shower parameters for several
different media, finding agreement with published results where
available.  We also report on radio pulse emission from the charged
particles in the shower, focusing on coherent emission at the
Cherenkov angle.  Previous work has focused on frequencies in the 100
MHz to 1 GHz range.  Surprisingly, we find that the coherence regime
extends up to tens of Ghz.  This may have substantial impact on future
radio-based neutrino detection experiments as well as any test beam
experiment which seeks to measure coherent Cherenkov radiation from an
electromagnetic shower.  Our study is particularly important for the
RICE experiment at the South Pole.
\end{abstract}

\tableofcontents

\section{Introduction}

Ultra high energy (UHE) neutrinos can travel without scattering over
large distances.  These may prove to be useful cosmological and
astrophysical probes.  They also present themselves as candidate high
energy particles with which we can test the Standard Model of
electro-weak theory beyond the energy regime of current
accelerators. In an UHE electron-neutrino charged current interaction,
the neutrino gives most ($\approx 80\%$) of its energy to the
secondary electron, which can then initiate an electromagnetic cascade
or shower.  It was predicted that an electromagnetic shower generated
by a high energy primary could develop a charge excess which would
emit Cherenkov radiation coherently \cite{askaryan62}.  For ultra high
energy primaries the Cherenkov radiation would be coherent in the
radio region of the spectrum; this long wavelength radiation might
then be detected using radio antennas \cite{markov86, provorov95,
ralmc89}.  Given the small predicted flux of ultra high energy
neutrinos \cite{Stecker96, halzen97, rachen99}, a suitable experiment
to detect ultra high energy neutrinos using this method requires a
large, dense (and radio-transparent) volume. Antarctic ice is suitable
for this purpose. A detailed analysis of all aspects of such an
experiment was done by Frichter, Ralston and McKay (FRM) \cite{fmr96},
using a simulation developed by Zas, Halzen and Stanev (ZHS)
\cite{zhs92}.  FRM concluded that a modest array of optimally designed
antennas could detect many events per year. The sensitivity of radio
detection peaks above 1 PeV, which compliments an optical array such
as AMANDA \cite{price96}.

The Radio Ice Cherenkov Experiment (RICE) at the South pole
\cite{rice99} is designed as a prototype detector of ultra high energy
neutrinos with energy $\ge$PeV using this method.  One basic
requirement for such an experiment is a reliable Monte Carlo
simulation of the shower development, Cherenkov radiation, detector,
and data acquisition system.  One can also test the idea of coherent
Cherenkov emission at accelerator facilities by dumping a beam of
photons or electrons into a dense target like sand or salt or any
other suitable medium.  Such tests have begun with experiments at
Argonne and SLAC \cite{saltzberg01}.  A Monte Carlo simulation which
can be easily adapted to such a test beam experiment, where Fresnel
and possibly near zone radiation is important, and one that can
include hadron showers conveniently, is clearly desirable.  The ZHS
simulation, designed for electromagnetic showers and Fraunhoffer (far
zone) detection, has been a powerful tool.  However an expansion and
update with extensive testing, offering applications to test beam and
neutrino astronomy, is currently needed.

We have written a GEANT-based Monte Carlo simulation to study coherent
Cherenkov emission in ice, salt, or a beam dump.  GEANT 3.21
\cite{geantman} is a well known and widely used simulation and
detection Monte Carlo package in particle physics\footnote{Differences
between GEANT 3.21 and GEANT 4 are primarily at energies below the
threshold for emission of Cherenkov radiation, and therefore do not
affect the results presented here.}.  It allows access to all details
of the simulation such as controls of various processes, definition of
target and detector media, and a complete history of all events
simulated. GEANT can be used to simulate all dominant processes in 10
keV - 10 TeV energy range, although it has not been extensively
verified for energies $>$100 GeV, where the extrapolation of
well-known lower-energy electromagnetic cross-sections becomes large,
and other effects (LPM, e.g. \cite{landau53-1,landau53-2}) become
significant. For electron energies above 10 GeV, GEANT uses screened
Bethe-Heitler cross-section for {\it bremsstrahlung} together with the
Migdal corrections \cite{messel70,migdal56}.  The first Migdal
correction is important for energy $\ge$1 TeV, reducing the
cross-section. The second correction reduces the differential
cross-section for soft photon emission and is effective even at much
lower energies, in the 100 MeV - 1 GeV range.  The LPM effect in the
context of UHE electromagnetic shower development and radio emission
has been discussed in \cite{misaki89, ralmc90, muniz97}.

GEANT is used to simulate electromagnetic showers inside materials
from which we extract detailed track information including particle
type, coordinates, energy and interaction time.  From this track
information, we investigate shower properties like radiation length,
Moliere radius, critical energy and energy deposition in the
material. We also determine the resulting radio pulse using standard
electrodynamic calculations from charged particles' tracks and by
parametrizing the shower.  Ultimately we will consider hadronic shower
information and GEANT provides the flexibility to expand our analysis
to this case.  We note that physics results presented thus far by RICE
have neglected the hadronic shower contribution and are, in this
respect, conservative.

The organization of this report is as follows. In section two, we
discuss various aspects of electromagnetic showers and define
quantities which characterize the shower. We present results on the
shower structure from the Monte Carlo simulation in section three and
compare them with established values in standard materials such as
iron, lead and carbon. Our analysis includes the detailed breakdown of
the radiation-generating charge imbalance into energy bins, the direct
evaluation from energy considerations of Moliere radius and the
determination from $\dd E/\dd x$ of radiation and energy deposition of
the critical energy. We discuss shower-to-shower fluctuations by
parametrizing the showers in section four. In section five, we review
the theory behind coherent emission of an electromagnetic pulse from
the shower \cite{allan}. In section six, we calculate the electric
pulse from the shower using track information from GEANT. We summarize
our results in section seven, making a number of comparisons with ZHS,
and discuss future work.

\section{Shower Description}

When a high energy electron or photon hits a material target, an
electromagnetic shower develops longitudinally inside the material.
At the beginning of the shower, {\it bremsstrahlung} and {\it pair
production} are the dominant processes.  Due to these processes, the
number of particles increases exponentially and the shower is created.
Due to the dominance of {\it bremsstrahlung} as the primary energy
loss mechanism for high energy electrons, the population of photons
quickly dominates that of electrons or positrons.  The energy of the
initial particle is divided among the secondaries.  The exponential
production of particles is halted when the charged particles reach the
critical energy ($E_{\rm c}$).  This is the transition region where
{\it ionization loss} overtakes {\it radiation loss} as the most
important electron energy loss mechanism.  The particle population
also reaches its maximum (``shower max'') at this point.  After
reaching the critical energy, charged particles lose their energy
predominantly inelastically by {\it ionization}, resulting in a
subsequent decline in the number of particles in the shower as they
degrade in energy and range out in the medium.

{\it Multiple Coulomb scattering} is responsible for the transverse
spread of the shower.  The shower core is populated by the highest
energy particles.  There is a long tail composed mostly of the
coulomb- scattered low-energy component.

Other processes - {\it Compton}, {\it Moller} and {\it Bhabha} and
{\it positron annihilation} build up a net charge (more electrons than
positrons) in the shower as atomic electrons in the target medium are
swept up into the forward moving shower.

\subsection{\it Energy Loss Mechanisms}

Energy loss of an electron due to radiation is well approximated by
the Bethe-Heitler formula \cite{bethe34}
\be -\left(\frac{\dd E}{\dd x}\right)_{\rm rad} = E \frac{4 N_{\rm o}
Z^2 r_e^2} {137 A} \, {\rm ln} \left(\frac{183}{Z^{1/3}}\right)
\label{eq:bethe-heitler} \ee
where $E$ is the initial energy of the electron, $x$ is the distance
in g-cm$^{-2}$ units, $N_{\rm o}$ is Avogadro's number, $A$ is the
mass number of the nuclei of the medium, $Z$ is the atomic number of
the medium and $r_{\rm e}^2 = e^4/m^2 c^4$ is the classical electron
radius. For a composite medium like ice ($H_2O$), one has to calculate
the effective $Z$ and $A$ using the proportion by weight
method\footnote{see for example CONS110 section of \cite{geantman}.}.

The Bethe-Bloch formula \cite{bethe30,bloch33} gives the energy loss
due to ionization as
\be -\left(\frac{\dd E}{\dd x}\right)_{\rm ion} = 4 \pi N_{\rm o}
\frac{Z}{A} r_e^2 m c^2 \left[{\rm ln} \left(\frac{2 m v^2
\gamma^2}{I}\right)-1-\frac{\delta}{2} \right]
\label{eq:bethe-bloch} \ee 
where $I = 10\,Z$ eV and $\gamma$ is the Lorentz factor in usual
relativistic notation. At very high energy, one needs to take into
account the {\it density effect} as the medium becomes polarized. This
is taken care of by the following term inside the square bracket of
Eq. (\ref{eq:bethe-bloch}):
\be \delta/2 = {\rm ln} (\hbar \omega_{\rm p}/I)+{\rm ln} \beta
\gamma -1/2, \label{eq:densityeffect} \ee 
where $\omega_{\rm p}$ is the plasma frequency of the medium.

A rough expression for the critical energy, $E_{\rm c}$, can be found
by the ratio of the expressions for radiation and ionization
losses. The log term in Eq. (\ref{eq:bethe-heitler}) is roughly $4$
and the square bracket term in equation (\ref{eq:bethe-bloch}) is
approximately $11$ if we include polarization effects. The ratio is
then
\be E_{\rm c} \approx \frac{605}{Z} \;\; {\rm MeV}.
\label{eq:crit-energy} \ee 

An alternate definition of the critical energy is that of Rossi
\cite{rossi}, according to which, the critical energy ($E_{\rm c}$) is
the energy at which the ionization loss per radiation length is equal
to the electron energy. Critical energies determined from these two
definitions are compared in a subsequent section.

The radiation length ($X_{\rm o}$) is given by
\be \frac{1}{X_{\rm o}} = \frac{4 N_{\rm o} Z^2 r_e^2}{137 A}
\,{\rm ln} \, \left(\frac{183}{Z^{1/3}} \right)
\label{eq:rad-length1} \ee 
so that the approximate expression for radiation loss (Eq.
(\ref{eq:bethe-heitler})) becomes $-(\dd E/\dd x) = E/X_{\rm o}$ or
\be <\!E\!> = E_{\rm o} e^{-x/X_{\rm o}}. \label{eq:rad-length3} \ee
Eq. (\ref{eq:rad-length3}) serves as a definition of the radiation
length; $X_{\rm o}$ is the thickness of material needed to reduce the
mean energy of an electron to $1/e$ of its original through {\it
bremsstrahlung}.

\subsection{\it Longitudinal Profile}

Heitler \cite{heitler} developed a simplified model of an
electromagnetic shower according to which the initial electron, with
energy $E_{\rm o}$, radiates a photon, of energy $E_{\rm o}/2$, in the
1st radiation length. In the 2nd radiation length, the photon creates
an electron-positron pair and the previous electron emits another
photon; each particle now has energy $E_{\rm o}/4$. This process
continues until the critical energy $E_c$, defined in the previous
section, is reached, at which point the shower is defined to be at
maximum.  At any radiation length, $t$, the number of particles
(electron, positron and photon) is $N(t)=2^t = {\rm exp} \;(t\,{\rm
ln}\,2)$ and the energy per particle is $E(t)=E_{\rm o}/2^t$. Thus in
this model the shower maximum occurs at $t_{\rm max} = {\rm
ln}\,(E_{\rm o}/E_c)/{\rm ln}\,2$ and the total number of particles at
the maximum is $N_{\rm max} = E_{\rm o}/E_c$. We can also calculate
the integrated track length of the charged particles as $L=(2/3\,{\rm
ln}\,2)(E_{\rm o}/E_{\rm c})$.

Although this model is over-simplified in assuming equality of all the
primary cross-sections and in assuming that the shower is artificially
cut off, it nevertheless provides a good description of the
qualitative features of an electromagnetic cascade. Namely, it
predicts that the total track length of the charged particles scales
linearly, while the position of the shower maximum scales
logarithmically with the initial energy of the shower.  These are
important consistency checks for the electromagnetic showers generated
from any Monte Carlo.

An analytic method for realistic shower development was created by
Carlson and Oppenheimer \cite{oppenheimer37} and later extended by
Landau and Rumer \cite{landau38}. Analytic shower equations are solved
in two approximations, namely {\it approximation A} and {\it
approximation B} \cite{rossi}. One neglects the Compton effect and
ionization loss in {\it approximation A}. One includes a constant
energy loss per radiation length for all electrons and positrons in
{\it approximation B}, which is therefore expected to be a better
model of the data.  The derivation of the {\it integral spectra} or
number of particles (electrons and positrons) at different shower
depth\footnote{Shower depth $t={\rm depth}/X_{\rm o}$ is always given
in terms of the radiation length ($X_{\rm o}$).} $t$ in approximation
A and B can be found in references \cite{rossi, gaisser}.

Greisen first parametrized the longitudinal profile of a
photon-induced electromagnetic air shower \cite{greisen56}. The
Greisen parametrization (GP) is based on analytic shower theory in
approximation A.  The difference between the more realistic
approximation B and approximation A is that of a slowly increasing
function of the {\it age parameter} ($s$). The number of particles in
a given energy range increases when $s<1$, reaches a maximum when
$s=1$ and then declines when $s>1$.

Hillas \cite{hillas82} later modified Greisen's parametrization to fit
Monte Carlo simulation of 0.1-1 TeV photon-induced showers.  This
modified GP was later used by Fenyves et. al. \cite{fenyves88} and is
given for a single electron or photon of energy $E_{\rm o}$ as:
\be N^{(\pi)} (E_{\rm o},E,t) = \frac{0.31\, A(E)}{\sqrt{y}}\;{\rm
exp}[t_1(1-1.5\, {\rm ln}\, s_1)]
\label{eq:greisenpara} \ee 
where $t_1$ is the modified depth defined to be $t_1 = t + a_{\pi,
\gamma}(E)$ and $s_1$ is the modified age parameter defined to be $s_1
= 3 t_1/(t_1+2y)$.  Superscript $\pi$ denotes the total electrons and
positrons following the convention in the literature.  The parameter
$y$ called {\it lethargy} is defined to be $y={\rm ln}(E_{\rm
o}/E_{\rm c})$, where $E_{\rm c}$ is the critical energy.  One finds
the parameters $A(E)$, $a_{\pi}(E)$ for electron-induced and
$a_{\gamma}(E)$ for photon-induced showers by fitting Monte Carlo
simulations.

\subsection{\it Lateral Spread}

The transverse development of the shower is well described by a
quantity called {\it Moliere radius} ($R_{\rm M}$).  It is determined
by the average angular deflection per radiation length at the critical
energy ($E_{\rm c}$) due to multiple Coulomb scattering.  The average
deflection is given by \cite{nishimura}:
\be <\!\delta \theta^2\!>= \left(\frac{E_{\rm s}}{E}\right)^2 \delta t
\label{eq:nishimura} \ee
where $E_{\rm s} = \sqrt{4 \pi/\alpha}\,m_{\rm e} c^2 = 21.21$ MeV is
the scale energy.

A numerical estimate of $R_{\rm M}$ is obtained by considering the
fraction of energy that escapes transverse to the shower axis
\cite{nelson66,bathow70}:
\be \frac{U(r)}{E}=\frac{\int_0^{\infty}\int_r^{\infty} E(r,t){\rm
d}r{\rm d}t}{\int_0^{\infty}\int_0^{\infty} E(r,t){\rm d}r{\rm d}t}
\label{eq:molfrac} \ee
where energy ($E$) is expressed as a function of shower depth ($t$)
and radial distance $r$ from the shower axis. By definition, ninety
percent of the shower energy is contained inside a cylinder of radius
$R_{\rm M}$ centered on the shower axis. I.e., $r=R_{\rm M}$ when
$U(r)/E=0.1$ in Eq.  (\ref{eq:molfrac}).  Moliere radius is
independent of the energy of the shower and depends only on the
tracking medium in general.

In Rossi's definitions, the {\it Moliere radius} is related to the
critical energy ($E_{\rm c}$) and radiation length ($X_{\rm o}$) of
the material \cite{nelson66,bathow70} through the equation
\be R_{\rm M} = \frac{X_{\rm o} E_{\rm s}}{E_{\rm c}},
\label{eq:moliere} \ee 
which follows from Eq. (\ref{eq:nishimura}).

\section{Shower Simulations}

The target medium in our GEANT simulations, unless otherwise stated,
is defined to be an ice cube of side 1 kilometer. Given the molecular
composition, GEANT calculates the effective atomic number, $Z=7.2$ and
an effective mass $A=14.3$ for the compound ice. Other parameters like
radiation length, absorption length and cross-sections are calculated
automatically, once $A$ and $Z$ are specified.

GEANT gives all the details of particle tracking information like
interaction points, total energy, energy lost in interaction,
interaction time and so on. We used double precision for all the
variables in our output data files to minimize roundoff errors.  {\it
Unless stated otherwise, all particles are tracked down to total
energy of 0.611 MeV}, which is lower than the energy at which
particles are still relativistic and emit Cherenkov radiation.

\subsection{\it Radiation Length}

The radiation length given by Eq. (\ref{eq:rad-length1}) depends on
the atomic and mass numbers of the material. However employing the
definition given by Eq. (\ref{eq:rad-length3}), we can fit an
exponential to the average energy of the injected electron at
increasing depths. The average is taken from the data of a number of
showers generated by the Monte Carlo. Our ability to recover the input
value of $X_{\rm o}$ will serve as an internal consistency check to
ensure that we are tracking all the particles of interest, along with
their energies. 

\begin{figure}
\vskip 6.75cm
\center
\begin{picture}(0,0)
\includegraphics{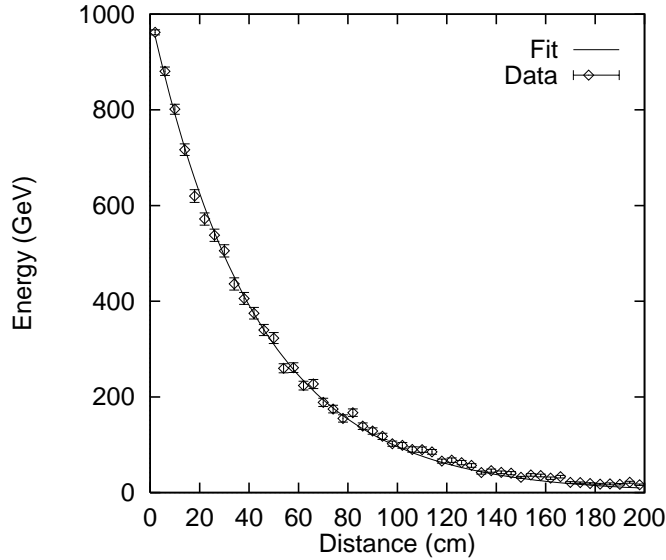}
\end{picture}
\vskip 0.5cm
\caption{\label{fig:rad}
         \small
Average energy of a 1 TeV electron injected in ice vs. distance in cm.
Monte Carlo data points are obtained from 500 electron tracks and the
errorbars correspond to standard error or $s/\surd N$, where $s$ is
the standard deviation and $N$ is the number of tracks (500).  The
solid line is the exponential fit as in Eq. (\ref{eq:rad-length3})
with $E_{\rm o} = 1$ TeV.  This fit gives a radiation length of $42.2
\pm 4.3$ cm.  A 100 MeV energy threshold was used in all 500
simulations.}
\end{figure}

We generated 500 electron tracks each with 1 TeV primary energy
$(E_{rm o})$ incident on the ice target and calculated energy loss due
to {\it bremsstrahlung}.  We tracked all the electrons down to energy
100 MeV, well above the value of the critical energy so that
bremsstrahlung is dominant.  We show the energy of the injected
electron, averaged over the 500 tracks, as a function of distance in
Fig. \ref{fig:rad}.  The errorbars correspond to standard error or
$s/\surd N$, where $s$ is the standard deviation and $N$ is the number
of tracks (500).  A least squares fit of Eq. (\ref{eq:rad-length3}) to
the Monte Carlo data keeping fixed $E_{\rm o} = 1$ TeV yields a
radiation length of $42.2 \pm 4.3 \;{\rm cm}$.  The confidence level
(CL) for the fit is 95.7\%.

Given the molecular composition, GEANT also calculates the medium's
radiation length from the standard formula.  For ice, GEANT calculates
$X_{\rm o}=38.8$ cm, which is roughly in agreement with the value we
extract by tracking bremsstrahlung photons.

\subsection{\it Moliere radius}

To calculate the Moliere radius, we construct an imaginary cylinder
centered on the shower axis (up to the physical length of the
shower). We add the energies, $U$, of all the tracks that leave the
cylinder without re-entering. By varying the radius of the cylinder we
obtain $R_{\rm M}$, which is the radius of the cylinder when the
fraction $U/E_{\rm o}$ is equal to 0.1.

\begin{figure}
\vskip 7.5cm
\center
\begin{picture}(0,0)
\includegraphics{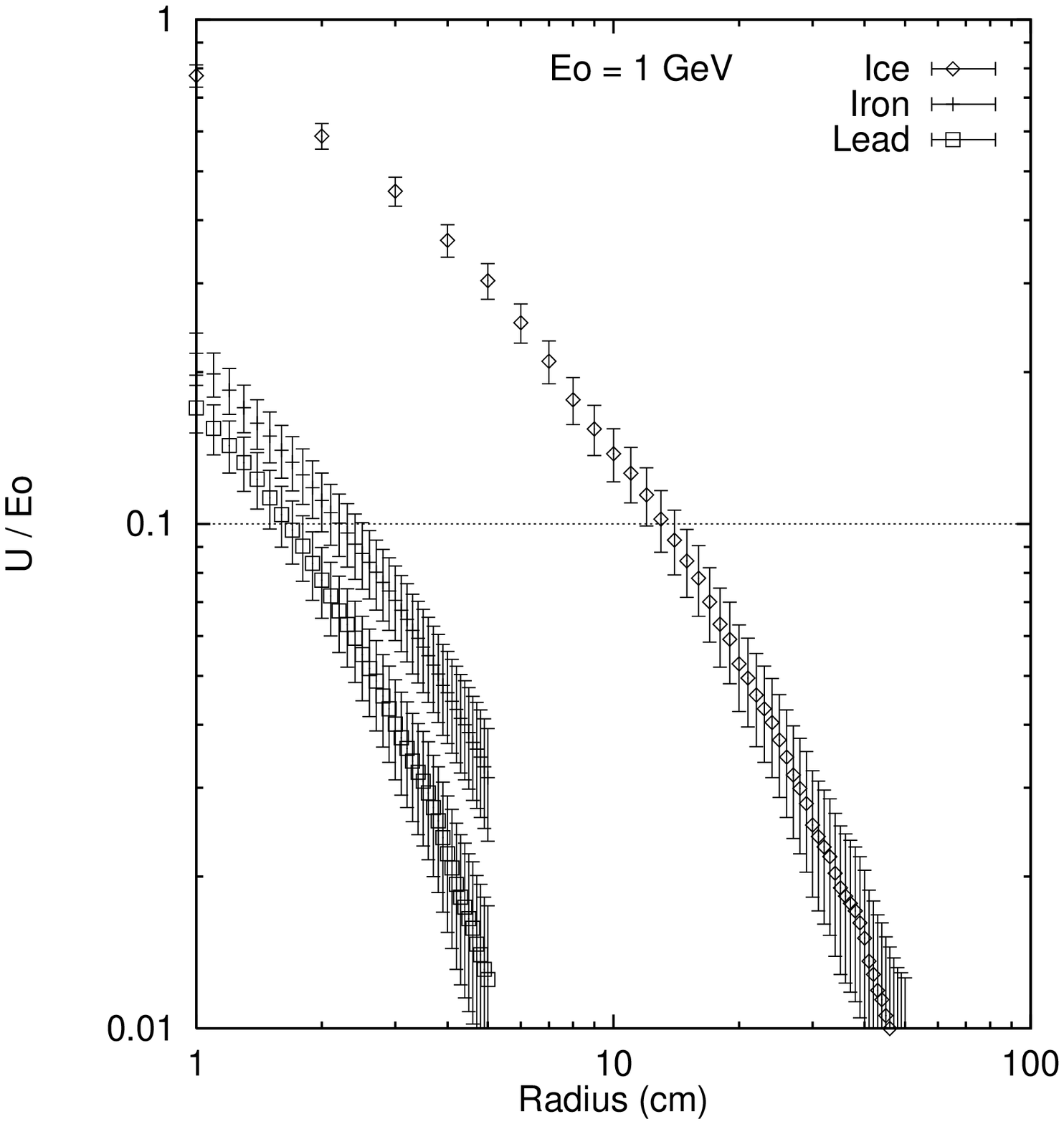}
\end{picture}
\begin{picture}(0,0)
\includegraphics{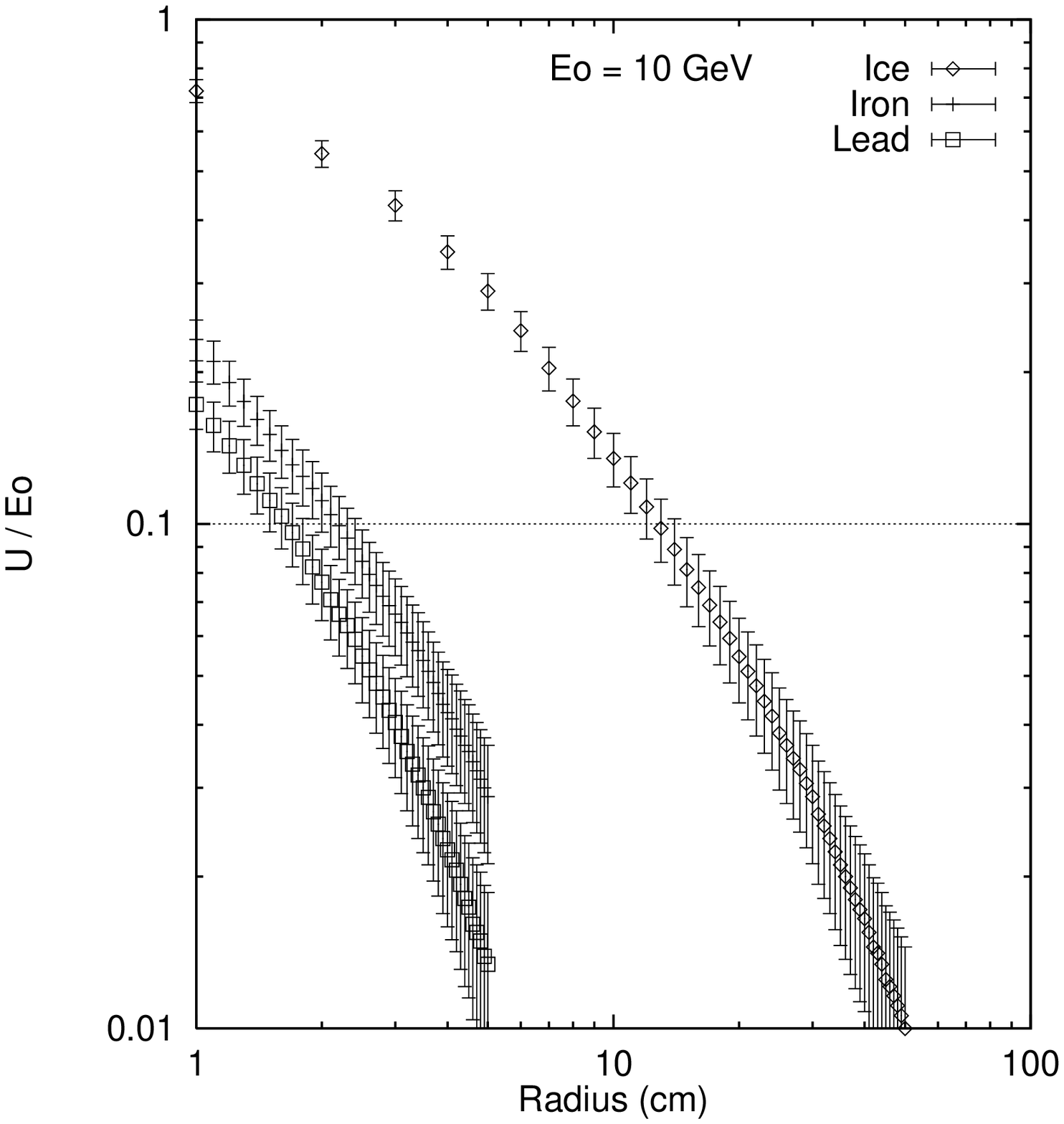}
\end{picture}
\vskip -0.5cm
\caption{\label{fig:molier}
         \small
Moliere radius corresponds to the transverse development of the
shower. When the fraction $U/E_{\rm o}$ (the ratio of total energy
inside an imaginary tube along the showers axis to the initial energy
of the shower) is 0.1 (the horizontal straight line), the
corresponding radius is by definition, the Moliere radius for that
material. It depends on the material and not on the energy of the
shower. Here we compare showers of energies 1 GeV (left plot) and 10
GeV (right plot) for each material; the Moliere radius is observed to
be independent of energy in this range.} \end{figure}

The Moliere radii ($R_{\rm M}$) for lead and iron are found to be 1.6
cm and 2.1 cm, respectively (see Fig. \ref{fig:molier}). The value for
lead is in agreement with experimental value of 1.6 cm
\cite{nelson66}. We did not find the corresponding value for iron in
the literature. A similar analysis for ice results in $R_{\rm M}^{\rm
ice}=13 \pm 1$ cm. We quote the standard error, defined as $s/\surd
N$, where $s$ is the standard deviation and $N$ is the number of
showers. It should be noted that the error bars are correlated from
bin to bin.

\subsection{\it Energy Loss in the Medium}

The signal from a Cherenkov type detector is proportional to the track
length, which is itself proportional to the energy deposition of the
shower particles in the medium.  Therefore, we study the ionization
loss per unit length in this section.

\begin{figure}
\vskip 6.cm
\center
\begin{picture}(0,0)
\includegraphics{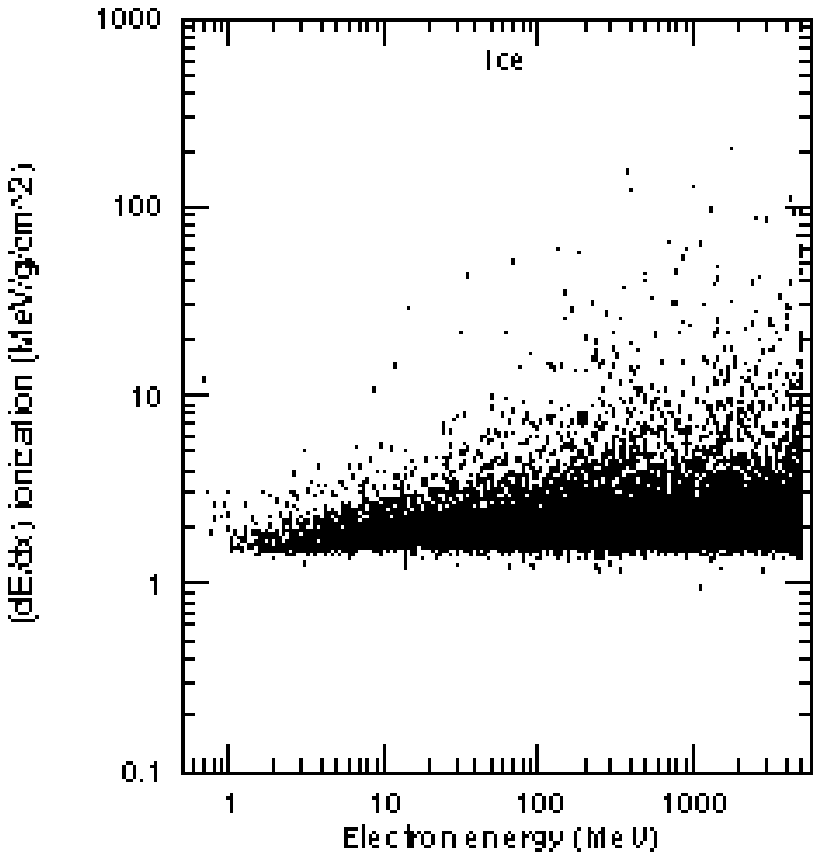}
\end{picture}
\begin{picture}(0,0)
\includegraphics{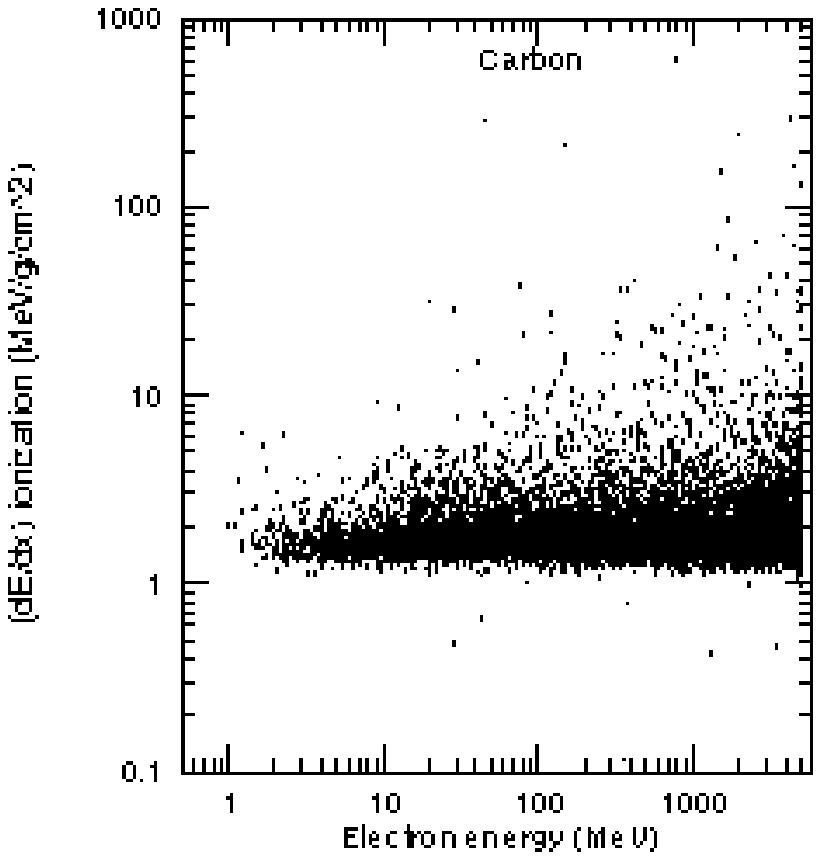}
\end{picture}
\caption{\label{fig:dedxdat}
          \small
Electron stopping power or $\dd E/\dd x$ ionization loss in ice (left
plot) and in carbon (right plot).  The dots correspond to energy lost
per unit length at each Monte Carlo step of 500 electron tracks each
with 5 GeV primary energy.  }
\end{figure}

To determine $\dd E/\dd x$ due to ionization from Monte Carlo, we
generated 500 separate 5 GeV electron tracks in ice and kept a record
of the rate at which energy was lost due to ionization.  The dots in
Fig. \ref{fig:dedxdat} (left plot) shows $\dd E/\dd x$ in ice for the
500 tracks at different Monte Carlo steps.  We also calculated $\dd
E/\dd x$ ionization loss in carbon, which has roughly the same atomic
weight as ice, as a consistency check. The result is shown in
Fig. \ref{fig:dedxdat} (right plot).

\begin{figure}
\vskip 7.cm
\center
\begin{picture}(0,0)
\includegraphics{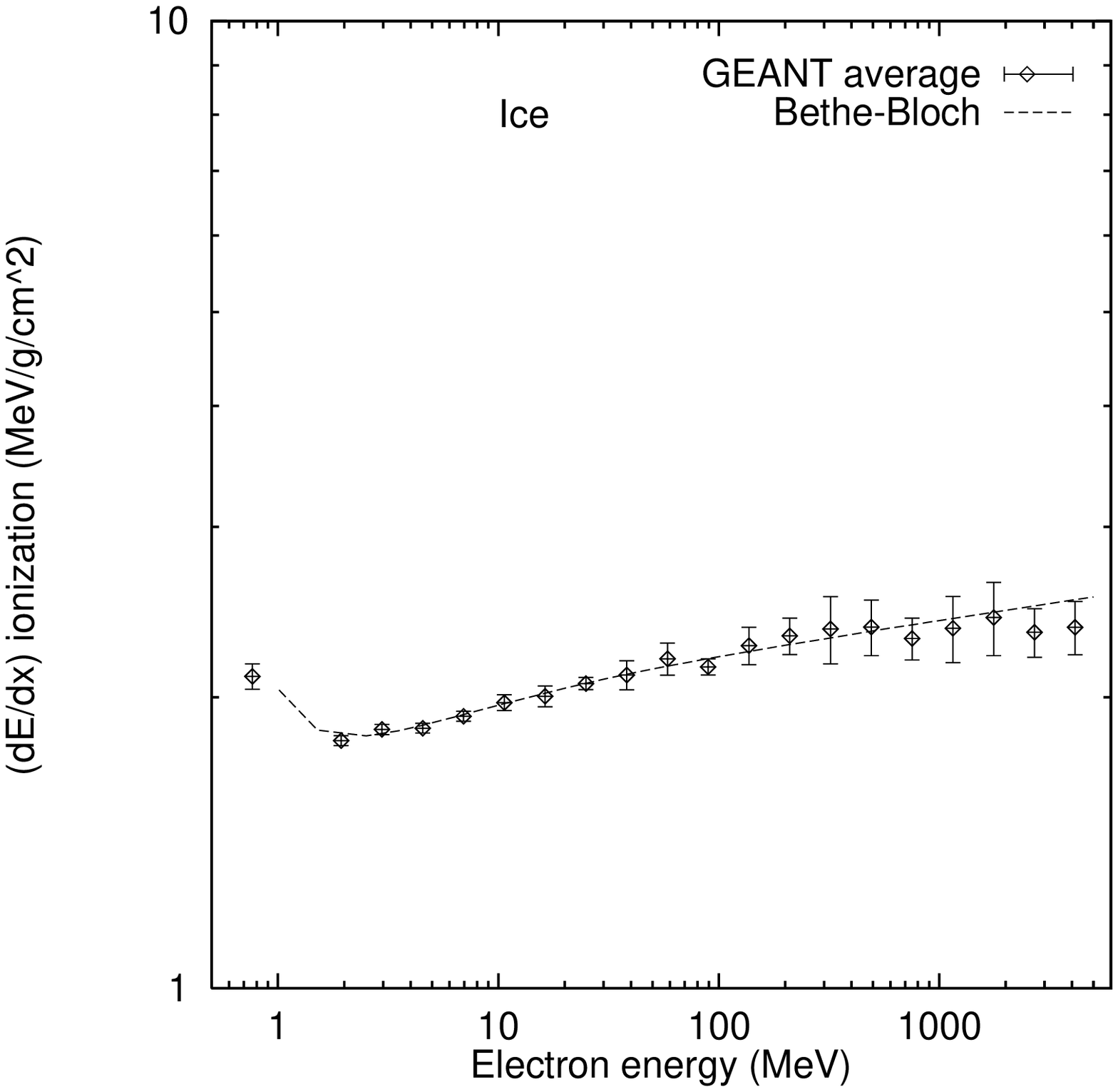}
\end{picture}
\begin{picture}(0,0)
\includegraphics{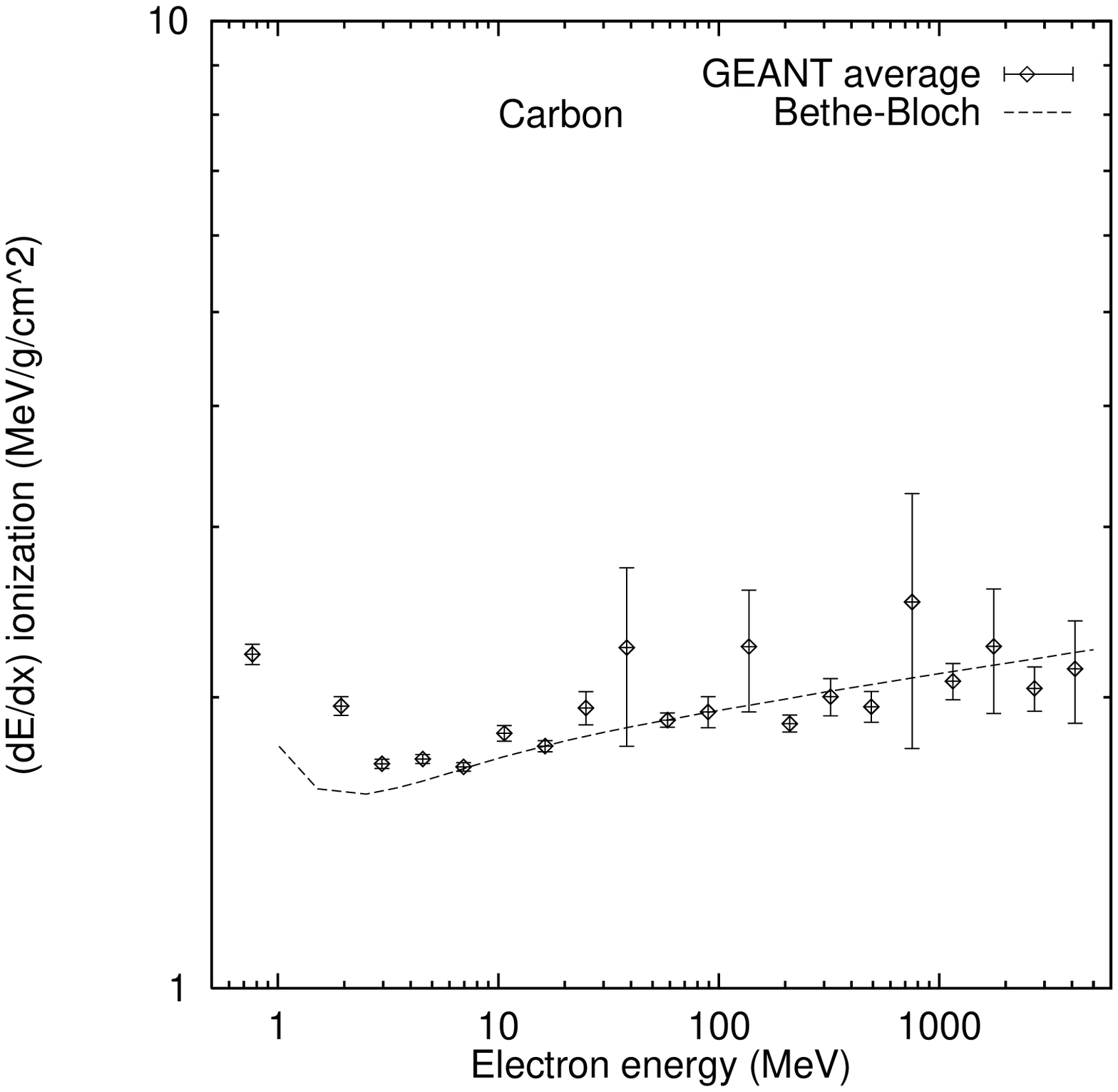}
\end{picture}
\vskip -0.5cm
\caption{\label{fig:dedxavg}
          \small
Average electron stopping power or $\dd E/\dd x$ ionization loss in
ice (left plot) and in carbon (right plot) calculated from GEANT data
in Fig. \ref{fig:dedxdat}. The error bars correspond to standard error
or $s/\surd{N}$, where $s$ is the standard deviation and $N$ is the
number of tracks (500 in this case). The solid lines are analytic
curves calculated from Eq.  (\ref{eq:bethe-bloch}). The agreement
between Monte Carlo and the analytic formula is reasonably good in the
relativistic rise region. }
\end{figure}

The average $\dd E/\dd x$ curve from GEANT simulation matches the
analytic approximation, Eq. (\ref{eq:bethe-bloch}) as shown in
Fig. \ref{fig:dedxavg}.  The average ionization loss in ice (left
plot) in the relativistic region is approximately 2.4 ${\rm MeV} {\rm
g}^{-1} {\rm cm}^{-2}$ and the average minimum ionization loss is
approximately 1.9 ${\rm MeV} {\rm g}^{-1} {\rm cm}^{-2}$.  We have
also calculated the average $\dd E/\dd x$ ionization loss in carbon
(right plot) as a consistency check.

\subsection{\it Critical Energy}

There are several methods to calculate the critical energy ($E_{\rm
c}$). The simplest is of course, the rough estimate in
Eq. (\ref{eq:crit-energy}). One can also obtain a value for $E_{\rm
c}$ in the Rossi definition from Eq. (\ref{eq:moliere}) after
supplying the value of the Moliere radius ($R_{\rm M}$) and the
radiation length ($X_{\rm o}$) found earlier. Yet another method is to
find the energy at which the $\dd E/\dd x$ curves due to ionization
and radiation losses intersect, which is the Monte Carlo equivalent of
the analytic expression in Eq.  (\ref{eq:crit-energy}).

First, we find the critical energy from the Rossi definition using
Eq. (\ref{eq:moliere}). The Moliere radius for ice was found to be
$R_{\rm M}^{\rm ice}=13 \pm 1$ cm and the radiation length, $X_{\rm
o}=42.2 \pm 4.3$ cm. The critical energy is then calculated from
Eq. (\ref{eq:moliere}) to be $E_{\rm c}^{\rm ice} = 68.8\pm 8.8$
MeV. Rossi \cite{rossi} quotes 65 MeV for the critical energy of
water, in agreement with our calculated value.  As a consistency
check, we also calculated the critical energy for lead using the
Moliere radii found from GEANT simulation (section 3.2). The result is
$E_{\rm c}^{\rm lead}= 7.4$ MeV, where we have used the radiation
length of lead to be 0.56 cm. This is in agreement with the
experimental result \cite{dovzhenko64}.

\begin{figure}
\vskip 6.75cm
\center
\begin{picture}(0,0)
\includegraphics{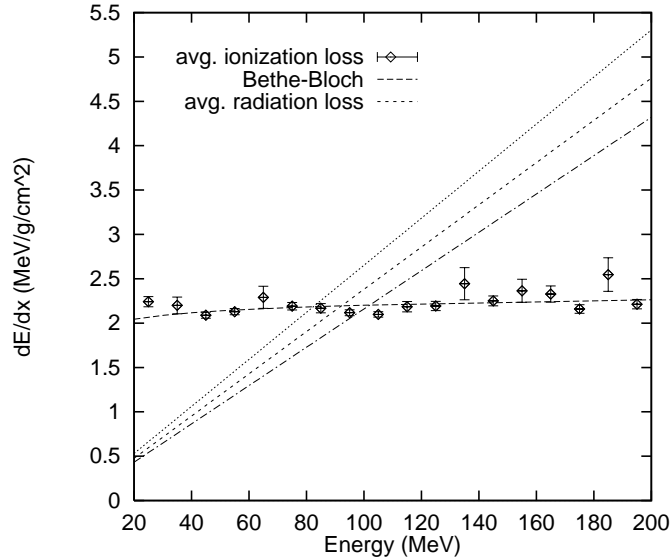}
\end{picture}
\vskip 0.5cm
\caption{\label{fig:critenergy}
         \small
Radiation and ionization energy loss vs. electron energy plot from
GEANT Monte Carlo simulation.  The diagonal straight lines correspond
to energy loss due to radiation with radiation length $X_{\rm o}=42.2
\pm 4.3$ cm, where the lowest curve corresponds to $X_{\rm
o}=42.2+4.3$ cm and the highest to $X_{\rm o}=42.2-4.3$ cm. The
average ionization energy loss data are plotted as obtained before
(see Fig. \ref{fig:dedxavg}). The error bars show the standard errors
on the $\dd E/\dd x$ points. The crossing point of the two curves
gives one definition of the critical energy. }
\end{figure}

Second, we calculate the energy loss due to radiation according to the
formula in Eq. (\ref{eq:rad-length3}) using the radiation length
$X_{\rm o} = 42.2 \pm 4.3$ cm we found before. We then find the energy
at which this line crosses the ionization loss curve (see
Fig. \ref{fig:critenergy}). The value for critical energy (the energy
at the crossing point of these two curves) found in this method is $90
\pm 9$ MeV, where the error includes both the standard error in the
$\dd E/\dd x$ points and the uncertainty in the radiation loss due to
the error in $X_{\rm o}$.  The approximate formula (Eq.
(\ref{eq:crit-energy})) gives $E_{\rm c} = 84$ MeV, close to what we
found from simulation.  The critical energy calculated in this way
gives a higher value than Rossi definition, as expected.

\subsection{\it Track Lengths}

If we consider all processes to be elastic except ionization, then an
estimate of the upper bound for the total track length is given by,
\be L = E_{\rm o}/\left(\frac{\dd E}{\dd x}\right)_{\rm ion}^{\rm min} 
\label{eq:track} \ee
where $(\dd E/\dd x)_{\rm ion}^{\rm min}$ is the minimum ionization
energy loss per unit length. This value is about $1.9\;{\rm MeV}{\rm
g}^{-1} {\rm cm}^{-2}$ for ice, (see Fig.  \ref{fig:dedxavg}) which
yields a maximum total track length of $\sim 1350$ radiation lengths
or 570 meters in ice for a 100 GeV shower. The actual total track
length is less than this value, since the energy loss due to
ionization increases in the relativistic region.

\begin{figure}
\vskip 6.75cm
\center
\begin{picture}(0,0)
\includegraphics{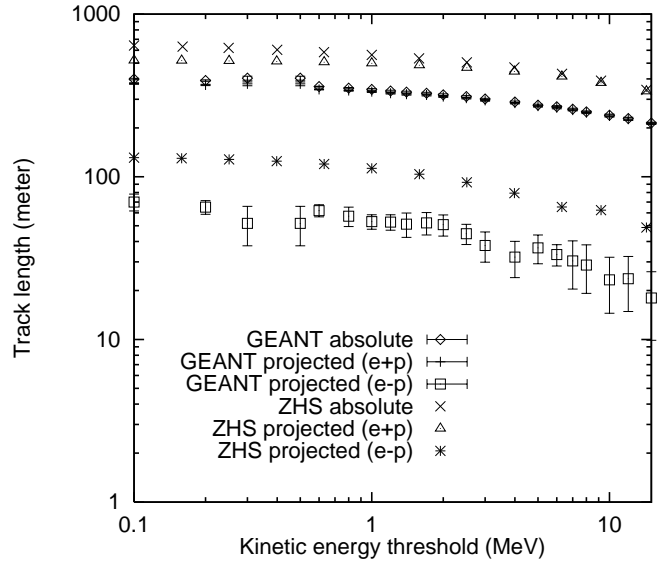}
\end{picture}
\vskip 0.5cm
\caption{\label{fig:trac1}
         \small
Total track lengths as functions of the kinetic energy threshold used
in the Monte Carlos. Here we show the total absolute track-lengths, the
total ($e+p$) track-lengths projected along the shower axis and the
total ($e-p$) track-lengths projected along the shower axis. The
analysis is done for a 100 GeV shower (averaged over 25 showers) using
the GEANT and ZHS Monte Carlo simulations. } 
\end{figure}

We plot different track lengths in Fig. \ref{fig:trac1} for a 100 GeV
shower (averaged over 25 showers) versus the kinetic energy threshold
used in the Monte Carlo to generate them. The total track length for
electrons and positrons together is denoted by total {\it absolute}
track length, the sum of electron and positron track-lengths projected
along the shower axis is denoted by total {\it projected ($e+p$)}
track length and the difference between the electron and positron
track lengths projected along the shower axis is denoted by total {\it
projected ($e-p$)} track length.  As we increase the threshold, so the
simulation neglects track-lengths from low energy particles, we expect
these total calculated track-lengths to decrease.

We have also compared our results to those obtained using the ZHS
Monte Carlo in Fig. \ref{fig:trac1}.  Although the qualitative
behavior of this scaling is the same for both Monte Carlos, ZHS track
lengths are consistently higher than GEANT track lengths by about
$50\%$. In particular, GEANT produces an absolute track length of 400
meter for 0.1 MeV kinetic energy threshold or 0.611 MeV total energy
threshold. ZHS on the other hand produces 650 meters of absolute track
length for the same threshold, which is significantly above our
estimated maximum of 570 meters, based on the GEANT generated 100 GeV
data shown in Fig. \ref{fig:trac1} and application of
Eq. (\ref{eq:track}).

The total track lengths increase with shower energy as more particles
are created. The track lengths are expected to increase linearly with
energy for a given a energy threshold used in the Monte Carlo. This
scaling is shown in Fig. \ref{fig:trac2} below.  Straight line fits to
the absolute track length and the projected ($e-p$) track length are
also plotted.  The slopes of those straight lines are 3.2 and 0.5
respectively.

\begin{figure}
\vskip 6.75cm
\center
\begin{picture}(0,0)
\includegraphics{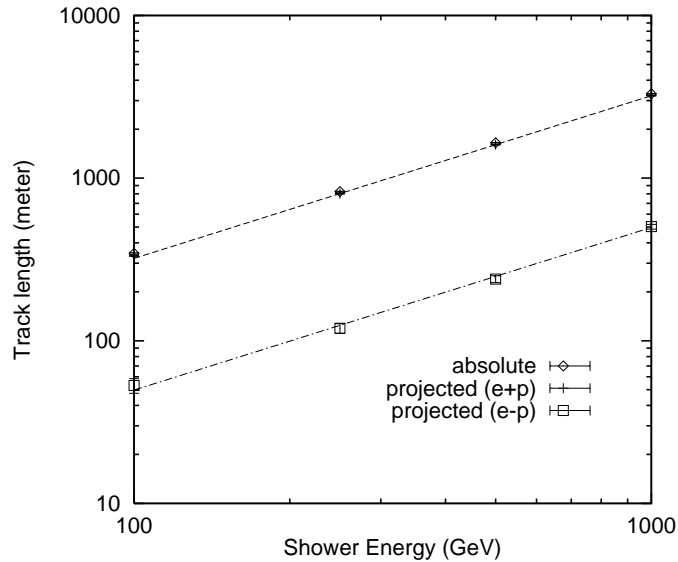}
\end{picture}
\vskip 0.5cm
\caption{\label{fig:trac2}
         \small
Total absolute track length, total projected ($e+p$) track length and
total projected ($e-p$) track length as functions of shower energy
from GEANT simulations. The linear scaling of the track lengths with
the shower energy is clearly observed. We used a kinetic energy
threshold of 1 MeV (see Fig. \ref{fig:trac1}) and averaged over 25
showers in each case.  }
\end{figure}

\subsection{\it Particle yield}

To generate a longitudinal profile of the shower, one counts the
number of particles crossing planes perpendicular to the shower axis
inside the ice.  We have calculated profiles for both the total number
of particles ($e+p$) and for the excess charge ($e-p$) in the shower.

\begin{figure}
\vskip 6.75cm
\center
\begin{picture}(0,0)
\includegraphics{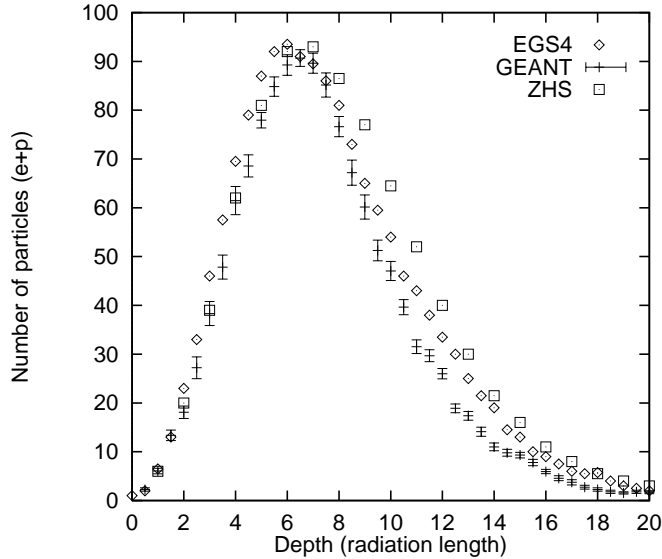}
\end{picture}
\vskip 0.5cm
\caption{\label{fig:depiron}
         \small
The longitudinal profiles of a 30 GeV shower (averaged over 50
showers) in iron simulated by GEANT. The longitudinal profile is
calculated by counting the number of electrons and positrons crossing
planes on the shower axis every half radiation length with total
energy greater than 1.5 MeV. We also plot the longitudinal profiles of
the same shower from EGS4 (from PDG 2000 \cite{pdg00}) and by ZHS
Monte Carlos. }
\end{figure}

As a consistency check of GEANT, we have simulated 30 GeV
electron-induced showers in iron and compared the profiles with the
modified Greisen parametrization, Eq. (\ref{eq:greisenpara}).  The
longitudinal profiles were obtained by adding the number of particles
with total energy greater than 1.5 MeV, crossing planes spaced
one-half radiation length apart, and perpendicular to the shower axis
(see Fig.  \ref{fig:depiron}).  The number of particles agrees
reasonably well with EGS4, an electromagnetic shower code developed at
SLAC which simulates the same shower \cite{pdg00}.  The total number
of particles from ZHS simulation is also shown in the plot.

\begin{figure}
\vskip 7.5cm
\center
\begin{picture}(0,0)
\includegraphics{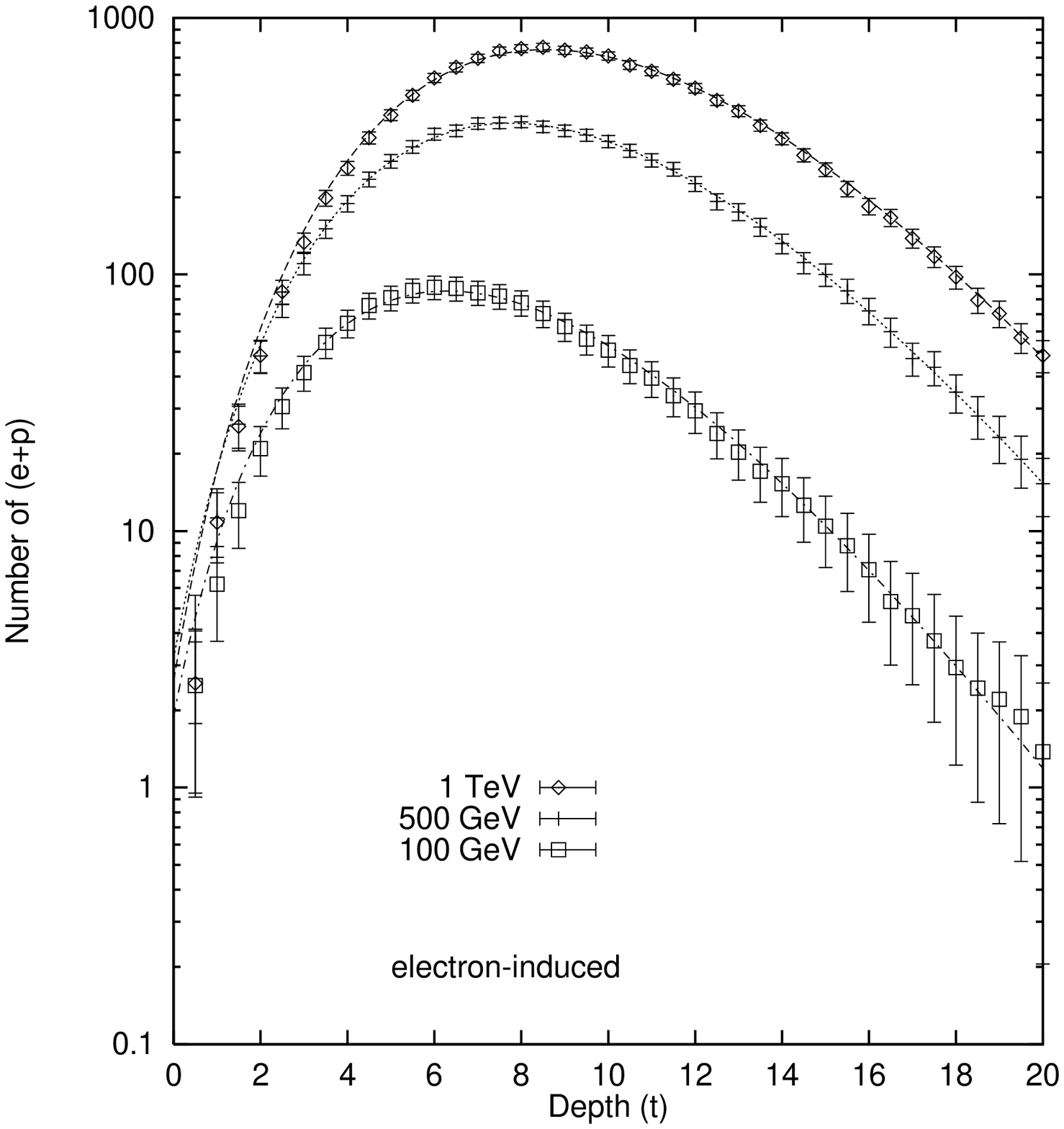}
\end{picture}
\begin{picture}(0,0)
\includegraphics{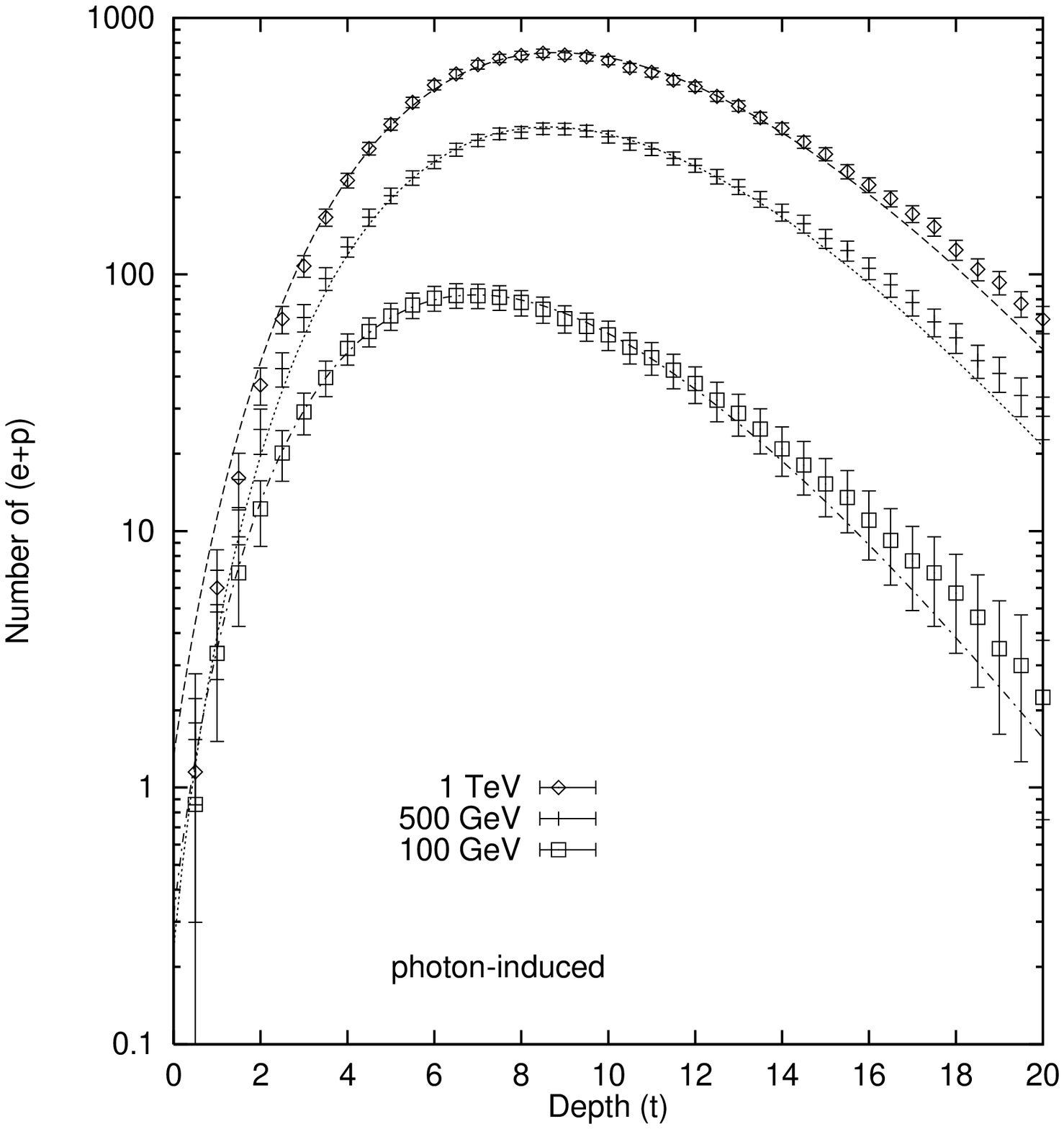}
\end{picture}
\vskip -0.5cm
\caption{\label{fig:greisenfits-eg}
         \small
Longitudinal profiles of electron-induced (left) and photon-induced
(right) showers in ice of primary energy 1 TeV, 500 GeV and 100 GeV,
averaged over 20, 50 and 100 showers respectively. The threshold
energy used is 5 MeV. The error-bars correspond to standard error. The
smooth curves are Greisen parametrization fits as given in
Eq. (\ref{eq:greisenpara}). The fitting parameters and the confidence
levels (CL) for the fits are given in Table 1. The critical energy for
ice used in the fits is 68.8 MeV.} 
\end{figure}

We have fitted the GEANT generated electromagnetic showers in ice by
the modified GP in Eq.  (\ref{eq:greisenpara}).  We used the value for
critical energy ($E_{\rm c}$) in ice to be 67.7 MeV.  The fitting
parameters and the confidence level (CL) of the fits for showers of
energy 100 GeV (averaged over 100 showers), 500 GeV (averaged over 50
showers) and 1 TeV (averaged over 20 showers) each with 5 MeV
threshold energy are given in Table 1 for both electron and photon
primaries.  The fits to longitudinal profiles are plotted in
Fig. \ref{fig:greisenfits-eg}.

\begin{center}
\begin{tabular}{c|cccc}
\multicolumn{5}{l}{Table 1: Greisen parameters for GEANT} \\
\multicolumn{5}{l}{showers in ice with 5 MeV threshold and} \\
\multicolumn{5}{l}{$E_{\rm c}=68.8$ MeV.} \\
\hline \hline
Primary & $E_o$ (GeV) & $A(E)$ & $a_{\pi,\;\gamma}$ & CL(\%) \\ \hline
        & 100  & 0.50 & 0.33 & 96.3 \\
$\gamma$ & 500 & 0.50 & 0.26 & 81.9 \\
        & 1000 & 0.50 & 0.76 & 71.8 \\ \hline
        & 100  & 0.52 & 0.99 & 95.6 \\
$e^-$   & 500  & 0.51 & 1.14 & 95.8 \\
        & 1000 & 0.52 & 1.01 & 88.4 \\
\hline \hline
\end{tabular}
\end{center} 

The parameter $A(E)$ gives over-all normalization for the number of
particles.  We see from Table 1 that the number of particles at the
shower maximum for a photon-induced shower ($A(E)=0.50$) is slightly
less than that of an electron-induced shower ($A(E)=0.52$).  The other
parameter $a_{\pi,\; \gamma}$ takes care of the shift of the whole
shower between a photon-induced ($a_{\gamma}$) and an electron-induced
($a_{\pi}$) shower.  This shift is approximately 0.7 radiation length.

\begin{figure}
\vskip 6.75cm
\center
\begin{picture}(0,0)
\includegraphics{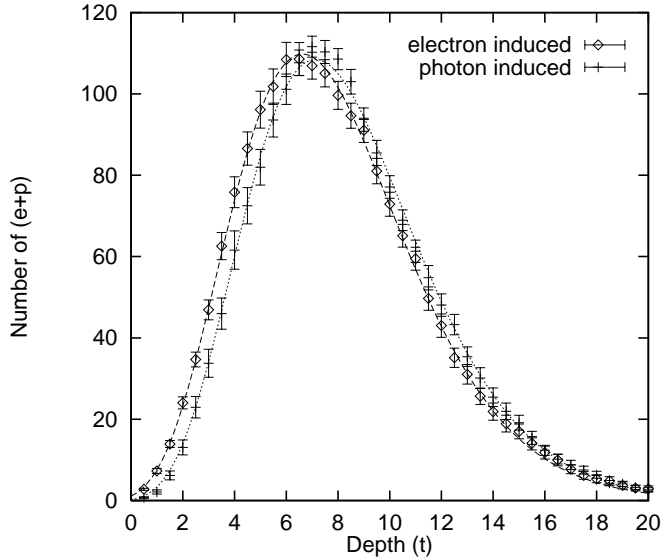}
\end{picture}
\vskip 0.5cm
\caption{\label{fig:dep100gev-greisen}
         \small
Greisen parametrization fits (smooth curves) to the longitudinal
profiles of 100 GeV electron and photon induced showers (each averaged
over 50 showers) with 0.611 MeV threshold energy from GEANT.  The
fitting parameters are given in Table 2.  The critical energy for ice
used in the fits is 68.8 MeV.}
\end{figure}

We have also fitted GEANT generated 100 GeV electron and photon
induced showers (each averaged over 50 showers) in ice with 0.611 MeV
threshold energy by GP given in Eq. (\ref{eq:greisenpara}).  The
fitting parameters and the CL for the fits are given in Table 2. The
fits with longitudinal profiles are plotted in Fig.
\ref{fig:dep100gev-greisen}. The critical energy in ice for the fits
is 68.8 MeV as before.

\begin{center}
\begin{tabular}{ccccc}
\multicolumn{5}{l}{Table 2: Greisen parameters for GEANT} \\
\multicolumn{5}{l}{showers in ice with 0.611 MeV threshold} \\
\multicolumn{5}{l}{and $E_{\rm c}=68.8$ MeV.} \\
\hline \hline
Primary  & $E_o$ (GeV) & $A(E)$ & $a_{\pi,\;\gamma}$ & CL(\%) \\ \hline
$\gamma$ & 100 & 0.65 & 0.16 & 85.6 \\
$e^-$    & 100 & 0.66 & 0.66 & 89.1 \\
\hline \hline
\end{tabular}
\end{center}

A three parameter fit to the GEANT generated 100 GeV electron-induced
shower in ice leaving the critical energy ($E_{\rm c}$) as a free
parameter along with $A(E)$ and $a_{\pi}$ gives $E_{\rm c} = 58.3 \pm
3.1$ MeV. This is within error bars of the critical energy ($E_{\rm
c}=68.8 \pm 8.8$ MeV) found from Moliere radius calculation.  The CL
for this fit is 97.7\%.

\begin{figure}
\vskip 6.75cm
\center
\begin{picture}(0,0)
\includegraphics{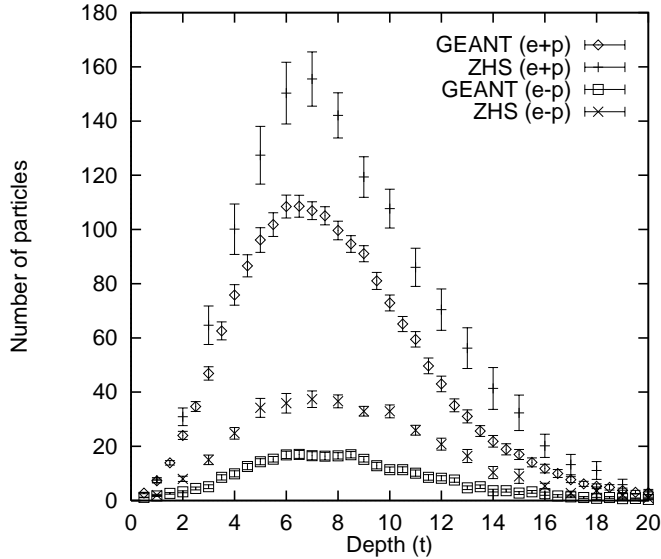}
\end{picture}
\vskip 0.5cm
\caption{\label{fig:depth2}
         \small
Comparison between shower profiles of a 100 GeV shower (averaged over
50 showers in each case) from GEANT and from ZHS Monte Carlos. The
error bars correspond to standard error, or $s/\surd N$, where $s$ is
the standard deviation and $N$ is the number of showers. The total
energy threshold used is 0.611 MeV in both cases.  The difference in
particle yield between the two Monte Carlos is $25-35\%$ and remains
the same at higher energy.  }
\end{figure}

A comparison of averages over 50 showers of 100 GeV each from GEANT to
the same from ZHS Monte Carlo (see Fig. \ref{fig:depth2}) shows about
a $25-35\%$ discrepancy for the total number of particles at the
shower max.  The percentage discrepancy between the two simulations
remains the same at higher energies.

The excess charge in the shower ($N(e-p)/N(e+p)$) is about 18\% at the
shower maximum, about twice Askaryan's original rough estimate
\cite{askaryan62}.

\begin{figure}
\vskip 6.75cm
\center
\begin{picture}(0,0)
\includegraphics{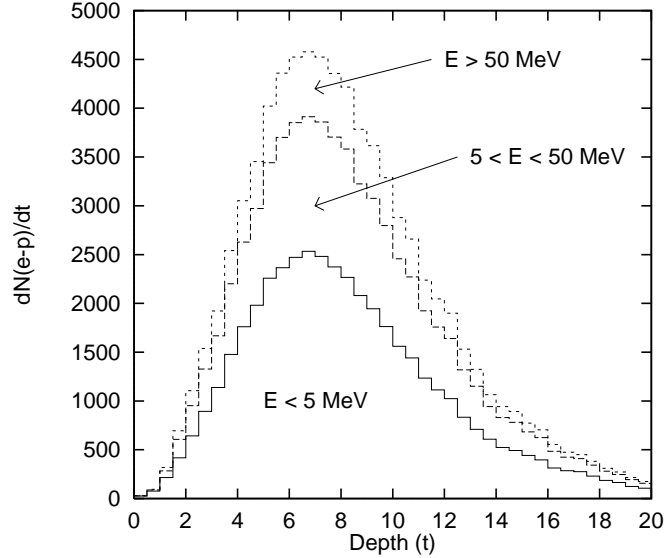}
\end{picture}
\vskip 0.5cm
\caption{\label{fig:depfrac}
         \small
Depth distribution of excess charge ($\dd N(e-p)/\dd t$) broken down
into 3 different energy bins: $E < 5$ MeV, $5 < E < 50$ MeV and $E >
50$ MeV. This figure shows that half of the excess electrons are
created with energy below 5 MeV. We used a 100 GeV shower (averaged
over 50 showers) with 0.611 MeV threshold to generate this plot.  }
\end{figure}

Fig. \ref{fig:depfrac} shows a longitudinal depth distribution of a
much different kind.  Unlike particles crossing planes transverse to
the shower axis, here we look at the distribution of {\it the complete
count of excess electrons} present in bins of one radiation length
along the shower axis. The figure shows the distribution of excess
electrons ($\dd N(e-p)/\dd t$) in a 100 GeV shower (averaged over 50
showers) with 0.611 MeV total energy threshold. The distribution is
broken down into different energy bins, showing that half of the
excess electrons have energy 5 MeV or lower.  This plot shows clearly
the important point that the bulk of the particles at shower maximum
have low energy and {\it contribute significantly to the Cherenkov
emission}.

\subsection{\it 1-D Model Study}

The comparisons in Figs. \ref{fig:depiron} \& \ref{fig:depth2} prompt
us to check the sensitivity of shower depth and particle number at
maximum to the cross-sections assumed in a simulation, since ZHS and
GEANT codes differ slightly in cross-sections for some processes (a
few percent).  To investigate this question in a setting that is
completely under our control, we have written a one dimensional Monte
Carlo that elaborates on Heitler's simple model to make it
realistic. We include Compton scattering, positron annihilation,
ionization loss, Moeller scattering, where interaction lengths are
chosen from exponential distributions. The default values for the
cross-sections are taken from GEANT.  We examine the effect of
changing cross-sections of various processes.  The total number of
particles ($e+p$ and $\gamma$) from the 1-D model scales linearly with
shower energy and the position of the shower maximum scales
logarithmically with energy as expected.

The effect of changing individual cross-sections (increasing and
decreasing by 25\%) are listed in Table 3 \& 4.  We also list the
total number of electrons and positrons ($N(e+p)$) at the maximum
depth ($t_{\rm max}^{e+p}$) and the total number of photons
($N(\gamma)$) at the maximum depth ($t_{\rm max}^{\gamma}$) for the
default values of the cross-sections (denoted by $\sigma_{\rm o}$).

\begin{center}
\begin{tabular}{l|ccccccc}
\multicolumn{8}{l}{Table 3: Results of increasing individual 
cross-section by 25\% in the} \\
\multicolumn{8}{l}{1-D Monte Carlo. The default setting (GEANT) is
denoted by $\sigma_{\rm o}$.} \\
\multicolumn{8}{l}{The cross-sections are changed one at a time from 
their default values.} \\
\hline \hline
Parameter & $\sigma_{\rm o}$  & Brem. & Pair & Comp. & $\delta$ 
& $\dd E/\dd x$ & Anni. \\ \hline
$t_{\rm max}^{e+p}$   & 6.9 & 7.2 & 6.2 & 6.9 & 5.6 & 6.4 & 6.7 \\  
$N(e+p)$        & 266 & 236 & 307 & 281 & 269 & 231 & 265 \\
$t_{\rm max}^{\gamma}$   & 6.9 & 7.0 & 5.9 & 6.4 & 6.4 & 6.7 & 6.4 \\
$N(\gamma)$     & 1418 & 1570 & 1495 & 1292 & 1396 & 1209 & 1416 \\
\hline \hline
\end{tabular}
\end{center}

\begin{center}
\begin{tabular}{l|ccccccc}
\multicolumn{8}{l}{Table 4: Results of decreasing individual 
cross-section by 25\% in the} \\
\multicolumn{8}{l}{1-D Monte Carlo.  The default setting (GEANT) is
denoted by $\sigma_{\rm o}$.} \\
\multicolumn{8}{l}{The cross-sections are changed one at a time from 
their default values.} \\
\hline \hline
Parameter       & $\sigma_{\rm o}$  & Brem. & Pair & Comp. & $\delta$ 
& $\dd E/\dd x$ & Anni. \\ \hline
$t_{\rm max}^{e+p}$   & 6.9 & 6.4 & 7.7 & 7.4 & 7.4 & 7.4 & 6.1 \\  
$N(e+p)$        & 266 & 310 & 212 & 249 & 298 & 298 & 271 \\
$t_{\rm max}^{\gamma}$   & 6.9 & 6.7 & 8.2 & 7.4 & 6.9 & 6.9 & 6.4 \\
$N(\gamma)$     & 1418 & 1173 & 1259 & 1516 & 1544 & 1544 & 1399 \\
\hline \hline
\end{tabular}
\end{center}

The individual cross-sections are changed one at a time from their
default values.  The changes in the values of $t_{\rm max}$ and $N$
are noticeable, but the percentage changes are much smaller than the
percentage changes in the cross-sections.  For example, comparing the
effect of increasing the bremsstrahlung cross section, listed in the
column under ``Brem.'' in Table 3, the depth and number at maximum
change by at most 4\% and 15\% respectively, compared to the default
values, listed under ``$\sigma_{\rm o}$, for a 25\% change in the
cross section.  It seems that the shower features are not especially
sensitive to precise values of an individual cross section. We have
not made an exhaustive study of the effects of changing combinations
of cross sections in all possible ways, but we have seen no indication
that effects would be large without pathological distortions of the
cross-sections.

\section{Shower Fluctuation}

Any determination of shower energy is intrinsically uncertain because
of the fluctuations from shower to shower for the same injected
energy.  The fluctuations must be quantified compactly so an efficient
and realistic energy uncertainty from this source can be assigned to
detected showers.  Fortunately, for UHE showers the huge numbers of
particles and fractionally smaller fluctuations make this less
problematic than for small showers.  We summarize an approach to the
fluctuation question in this section.

Conservation of total energy plays an important role in shower
development. The shape of a shower strongly depends on the position
and energy of the first hard bremsstrahlung - the later the first hard
bremsstrahlung event, the deeper the shower maximum is.  However, for
any shower, the primary energy (i.e., the total energy of the shower)
qualitatively dictates the energy loss profile and the particle yield
with depth.

It is common practice to describe the mean longitudinal profile of an
electromagnetic shower by a Gamma distribution which is similar in
shape to the Greisen parametrization. The Gamma distribution
normalized to unity is given by
\be f(t;a,b)=b \, \frac{(bt)^{a-1}e^{-bt}}{\Gamma(a)}.
\label{eq:gama} \ee 

However, Grindhammer et. al. \cite{grindhammer} have shown that the
fluctuations of the parameters $a$ and $b$ from an average profile do
not necessarily follow the individual shower fluctuations. It is more
reasonable to fit individual profiles by the Gamma distribution. One
can then fit a 2-dimensional Gaussian distribution to the parameter
set $\{a, b\}$ thus obtained and study shower fluctuation by studying
these parameters. The correlation between these parameters $a$ and $b$
can be expressed as:
\be \rho={\rm Covariance} \;(a,b)/\sigma_a \sigma_b,
\label{eq:correlation} \ee 
where $\sigma_a$ and $\sigma_b$ are standard deviations of the
parameters $a$ and $b$ respectively and ${\rm Covariance} \;(a,b)$ is
the covariance matrix. The correlation $\rho$ is roughly independent
of the energy of the shower.

\subsection{\it Fit to Simulations}

\begin{figure}
\vskip 6.65cm
\center
\begin{picture}(0,0)
\includegraphics{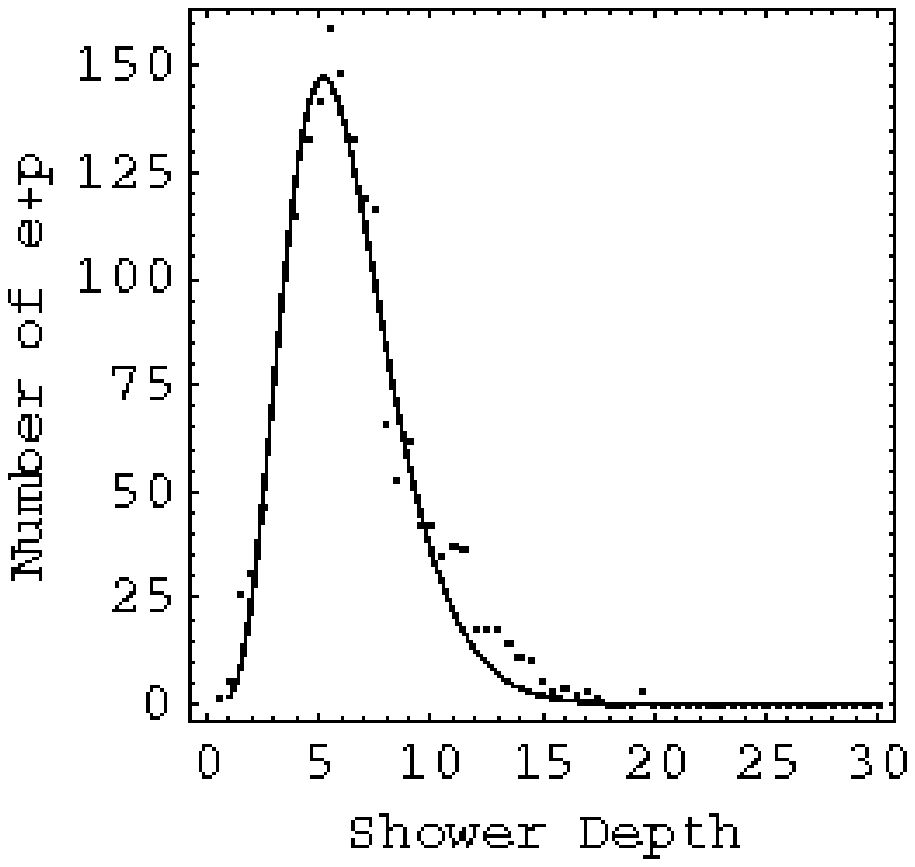}
\end{picture}
\begin{picture}(0,0)
\includegraphics{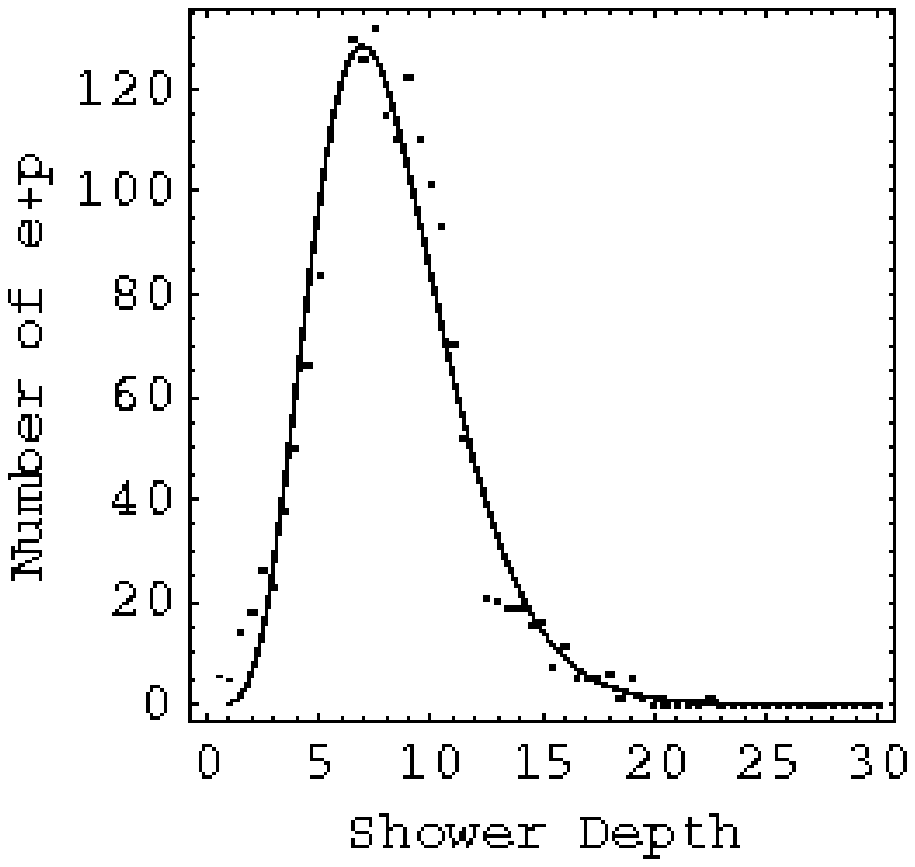}
\end{picture}
\caption{\label{fig:individual-profiles}
         \small
Individual shower profiles of 100 GeV showers from GEANT with 0.611
MeV threshold energy.  The solid curves are Gamma function fits to the
profiles.  The parameters $a$ and $b$ in Eq. (\ref{eq:gama}) are
obtained from the fits.}
\end{figure}

We have fitted the Gamma distribution Eq. (\ref{eq:gama}) to 50 GEANT
generated individual shower profiles in ice\footnote{particle number
($e+p$) normalized to 1.}. The energy of each shower is 100 GeV with
0.611 MeV threshold energy. The parameters $a$ and $b$ are extracted
from the Gamma function fits. Two such fits to individual shower
profiles are plotted in Fig.  \ref{fig:individual-profiles}.

\begin{figure}
\vskip 6.5cm
\center
\begin{picture}(0,0)
\includegraphics{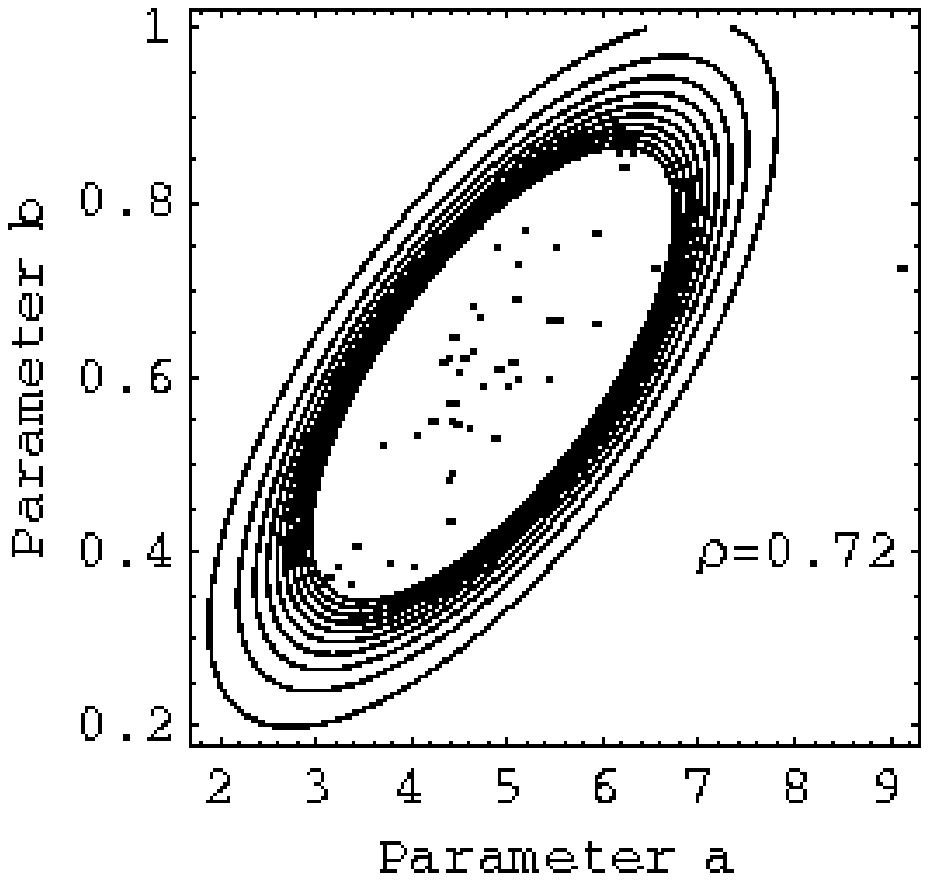}
\end{picture}
\begin{picture}(0,0)
\includegraphics{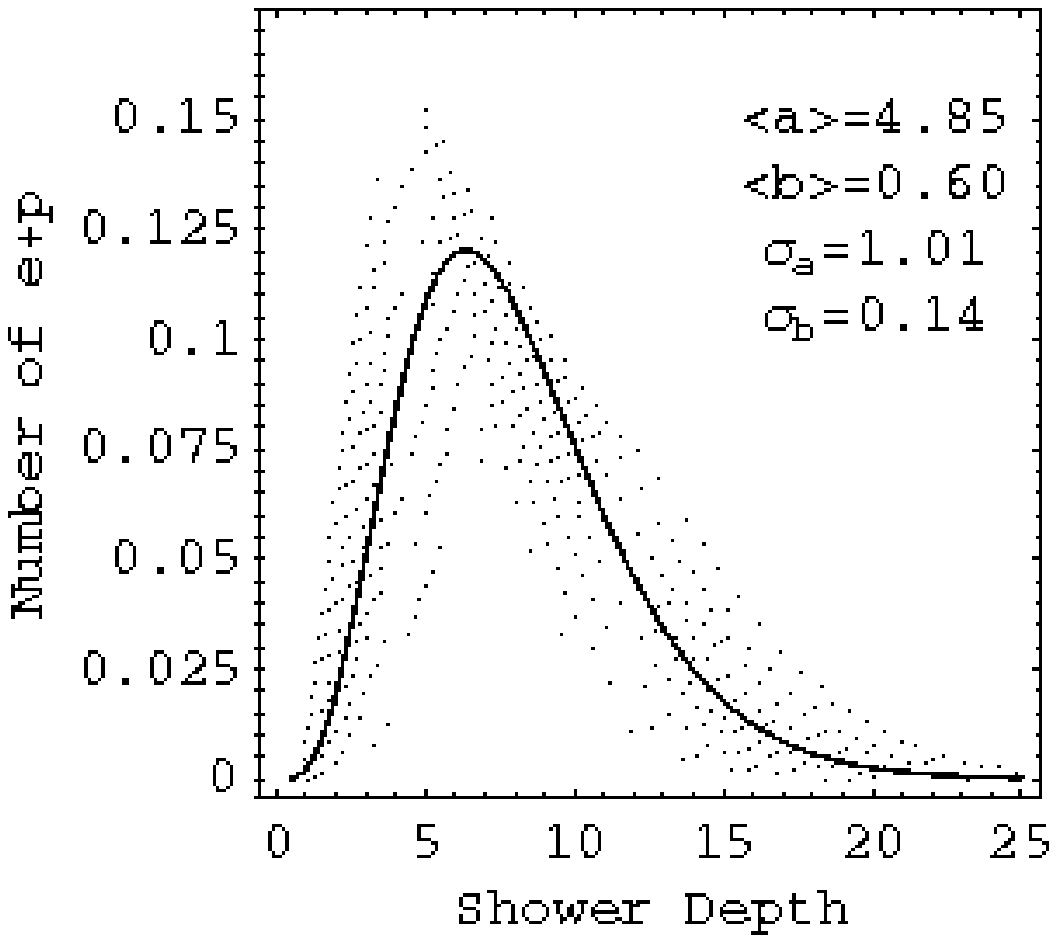}
\end{picture}
\vskip 0.5cm
\caption{\label{fig:shower-fluc}
         \small
Scatter plot (left plot) of the parameters $a$ and $b$ from the Gamma
distribution fit to 50 individual shower profiles each with energy 100
GeV and 0.611 MeV threshold.  The contours are from a 2-dimensional
Gaussian distribution with mean ($a$, $b$) and ($\sigma_a$,
$\sigma_b$) obtained from the data set ${a, b}$ generated as described
in the text.  Shower fluctuations (right plot) due to variation of the
parameters $a$ and $b$ within a standard deviation. The dark solid
curve is the profile with mean values of $a$ and $b$. All particle
numbers are normalized to 1.}
\end{figure}

Fig. \ref{fig:shower-fluc}a shows the scatter plot of the parameters
$a$ and $b$ obtained by fitting 50 individual showers.  We also show
the contours of the 2-dimensional Gaussian distribution fit to the set
of parameters $\{a,b\}$.  This plot is similar in shape to the one
obtained by Grindhammer using GEANT \cite{grindhammer}.  The
correlation coefficient $\rho=0.72$ we found is close to $0.73$ as
found by Grindhammer et. al.

Fig. \ref{fig:shower-fluc}b shows shower-to-shower fluctuations as we
vary the parameters $a$ and $b$ within a standard deviation
($\sigma_a$ and $\sigma_b$) of their mean values ($<\!a\!>$ and
$<\!b\!>$). The dark solid curve corresponds to the shower with mean
values of the parameters $a$ and $b$.  As can be seen, the shower
maximum can vary as much as $\sim 50\%$.  As remarked above, the
variations become statistically less significant as the energy rises
to the multi-PeV range, where UHE neutrino detectors such as RICE are
sensitive.  The uncertainties from fluctuations, though requiring
study, are not expected to be a major part of the uncertainties in
event locations and energies in any case.

\section{Electromagnetic Pulse Theory}

\subsubsection*{Overview}

Given a current density, one calculates the electric field by
straightforward application of Maxwell's equations.  The most useful
form of the electric field is the complex vector $\vec E_{\omega}(\vec
x)$, depending on three-dimensional coordinate $\vec x$ and angular
frequency $\omega$. Before getting into the details of the radiation
field calculation, we review its general features and several subtle
points involving coherence, and near and far field limits.

The energy radiated per unit frequency interval per unit solid angle
is proportional to $|E_{\omega}|^{2}.$ In an asymptotically far field
limit $R \ra\infty$, the radiation field $E_{\omega} \propto 1/R$,
where R is the distance from the shower to the field point.  Because
the quantity $E_{\omega}$ has dimensions of mass in particle physics
units and depends linearly on the track length $L$ (reviewed below),
one expects the frequency dependence $ E_{\omega} \propto L\omega /R$
on dimensional grounds. This linear dependence on $\omega $ breaks
down at high frequencies, when wavelengths are smaller than the
typical dimensions of the shower and create a coherence cutoff. The
purpose of this section is to develop the concepts and formalism
needed to address this and other issues.

The asymptotically far field is not a feature of Cherenkov physics as
commonly presented in texts or applied in particle physics detectors.
In particle physics applications, the track length is comparable to
the distance to the observation point $R$.  Textbook treatments take
the limit that the track length is infinite, \cite{jackson, low,
landau} in which case the electric field $E_{\omega} \sim
\sqrt{\omega}/ \sqrt{\rho}$, where $\rho$ is the {\it cylindrical}
radial coordinate.  In this situation, the radiation intensity is
proportional to $\omega$, a familar feature of laboratory Cherenkov
radiation.  The cylindrical symmetry dictates this special frequency
and radial dependence.  The transition from cylindrical configuration
to spherical configuration is associated with the terms ``Fresnel and
Fraunhoffer'' zones.  (Conditions of ``far fields'' $R \gg \lambda$,
$\rho \gg \lambda$ are separate and assumed throughout; criteria
separating the zones are given below.)  The conditions of the RICE
experiment are such that spherical symmetry is a good approximation
for the most distant events, $R \sim 1$ km, while Fresnel zone effects
start to become noticeable for the nearest likely events, $R \sim 100$
m.  Consequently both are reviewed.

There are two methods to calculate the electric field from the charged
particles in a shower.  One method calculates the electric field from
each charged track and adds them by superposition.  The other method
parametrizes the shower's current and calculates electric fields
analytically.  Direct track-by-track calculation and superposition are
presently limited to the Fraunhoffer zone.  The result is simple, and
depends only on the angle to the observation point.  The analytic
method works in either zone, and gives a compact way to take a few
parameters characterizing the shower and calculate the field.  Of
course one must have a good parameterization for this to be a good
approximation.

By using both the analytic and numerical methods, with their different
strengths, we are able to characterize the coherence structure of
showers in considerably more detail than previous work.  The coherence
extends to much higher frequencies than previously thought.  This is a
very important point, confirmed by two independent methods, and clears
up a misconception that coherence should cut off at about 1 GHz.

\subsection{\it Fraunhoffer Limit}

The Fraunhoffer limit is appropriate to most of the RICE sensitive
volume.  It forms the basis for our numerical, track-by-track
computation of the field produced by a shower at antenna sites remote
from the shower.  In this subsection, we derive the expression for the
field produced by an individual track, which forms the basis for the
calculation of the full field calculated from a whole shower. In the
next subsection, we outline the method of parametrizing the shower as
an effective current and calculating the field directly from that
current.

The power at time $t$, position ${\vec x}$, radiated per unit solid
angle by a moving electric charge is given, in Gaussian units, by
\cite{jackson}
\be \frac{\dd P(t)}{\dd \Omega} = \frac{c}{4\pi}|R{\vec
E}({\vec x}, t)_{\rm ret}|^2, \label{eq:power} \ee
evaluated at the retarded time $t=t'+R(t')/c$.  (For now we calculate
effects in vacuum; shortly we will supply the factors for the effects
of a medium.)  We use the Fourier transformed variables
\ba R{\vec E}_{\omega} &=& \frac{1}{\sqrt{2 \pi}}
\int_{-\infty}^{\infty}R{\vec E}(t) e^{i\omega t} \dd t \; ;
\label{eq:fourier1} \\ R{\vec E}(t) &=& \frac{1}{\sqrt{2 \pi}}
\int_{-\infty}^{\infty}R{\vec E}_{\omega} e^{-i\omega t} \dd
\omega \; . \label{eq:fourier2} \ea
The energy radiated {\it per unit frequency interval} per solid angle
is then given by
\be \frac{\dd^2 I}{\dd \omega \dd \Omega} =
\frac{c}{4\pi}|RE_{\omega}|^{2}
\label{eq:energy}. \ee
We will require the frequency dependence of the fields, so we work
with the Fourier transformed fields below.

\begin{figure}
\vskip 6.cm
\center \begin{picture}(0,0)
\includegraphics{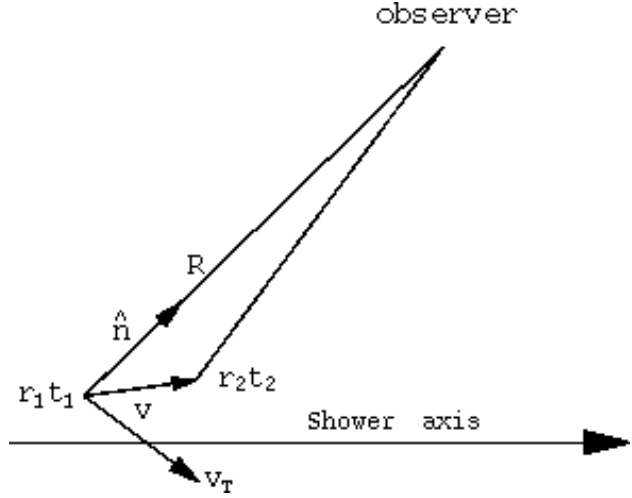}
\end{picture}
\vskip 0.5cm
\caption{\label{fig:diag}
\small Geometry for calculating electromagnetic fields from a single
track segment.  $(r_1, t_1)$ and $(r_2, t_2)$ are the starting and
ending position and time of the segment along which the particle moves
with velocity $\vec v$.}
\end{figure}

The expression for the radiation field from a point source is
conventionally defined as the electric field term linear in the
acceleration $\dot{\vec \beta}$: 
\be {\vec E}({\vec x},t)=\frac{q}{c}\left[\frac{{\hat n}\times
\{({\hat n}-{\vec \beta}) \times {\dot {\vec \beta}}\}}{(1-{\vec
\beta}\cdot{\hat n})^3 R}\right]_{\rm ret} \label{eq:electric} \ee
where $\vec{\beta}$ is the velocity of the particle, $\hat n$ is the
direction of the observer and $R$ is the distance from the track to
the observation point (see Fig. \ref{fig:diag}). The factor $1/R$ that
accompanies the $\vec{\beta}$ factor is the other trademark of the
radiation field.  As is the case for the term that comes from the
boosted Coulomb field, which has no explicit acceleration dependence,
Eq. (\ref{eq:electric}) is singular at the Cherenkov angle in a medium
with real index of refraction greater than 1.

Combining Eq. (\ref{eq:fourier1}) and Eq. (\ref{eq:electric}) we have
\be {\vec E}_{\omega}({\vec x})=\sqrt{\frac{q^2}{8\pi^2 c}}
\int_{-\infty}^{\infty} e^{i\omega (t'+R(t')/c)}\left[\frac{{\hat
n}\times \{({\hat n}-{\vec \beta}) \times {\dot {\vec
\beta}}\}}{(1-{\vec \beta}\cdot{\hat n})^2 R}\right] \dd t'.
\label{eq:fourier3} \ee
At distances large compared to the range of motion of the source,
${\hat n}$ is approximately constant and 
\ba R(t') \approx |\vec x|-{\hat n}\cdot{\vec r}(t') \label{frapprox} .
\ea
This is the {\it Fraunhoffer approximation}: the error in the phase 
$\omega |\vec x-\vec x'|/c$ must be kept small compared to
$2 \pi$.  Conditions for use of this approximation are discussed in
the next subsection\footnote{Analysis of the effects of keeping the
next order in the expansion of the phase shows that significant
deviations from the Fraunhoffer result appear at distances where the
Fresnel zone sets in \cite{florian}, as discussed from a general point
of view below \cite{roman00}.} (see Eq. (\ref{eq:ral4})).

After integrating Eq. ($\ref{eq:fourier3}$) by parts, and using the
boundary conditions to set the end point contributions to zero, one
finds \cite{jackson}
\be
R{\vec E}_{\omega}({\vec x})\approx -i\omega\sqrt{\frac{q^2}{8\pi^2
c}}e^{i\omega R/c} \int_{-\infty}^{\infty} e^{i\omega (t-{\hat
n}\cdot{\vec r}/c)}[{\hat n}\times({\hat n}\times{\vec \beta})]{\rm
d}t ,
\label{eq:fourier4}
\ee
where $R\equiv |\vec x|$.  We will later apply this track-by-track
expession to segments over which ${\vec \beta}$ is constant with time
from $t_1<t<t_2$.  On each segment ${\vec r}={\vec r}_1+ c{\vec
\beta}(t-t_1)$.  We insert the result into Eq.  (\ref{eq:fourier4})
and obtain
\be R{\vec E}_{\omega}({\vec x})=\frac{1}{\sqrt{2
\pi}}\left(\frac{q}{c}\right) e^{ikR}
e^{i\omega(t_1-{\hat n}\cdot{\vec r}_1/c)} {\vec \beta}_{\perp}
\frac{(e^{i\omega \delta t (1-{\hat n}\cdot{\vec \beta})}-1)}{1-{\hat
n}\cdot{\vec \beta}} \label{eq:pulse1} \ee
where ${\hat n}\times({\hat n}\times{\vec \beta})=-{\vec
\beta}_{\perp}$, and $\delta t=t_2-t_1$.

In a medium we replace $c\rightarrow c/n = c/\sqrt{\epsilon\mu}$,
where $n=n(\omega)$ is the refractive index of the material and
$\epsilon =\epsilon(\omega) $, the dielectric constant.  We also
replace ${\vec E}\rightarrow {\vec D}=\epsilon {\vec E}$ and ${\vec
\beta} \rightarrow {\vec v}\frac{n}{c}=n{\vec \beta}$ everywhere, to
get
\be R{\vec E}_{\omega}({\vec x})=\frac{1}{\sqrt{2 \pi}} \left(\frac{\mu_r
q}{c^2}\right) e^{ikR} e^{i\omega(t_1-\frac{n}{c}{\hat n}\cdot{\vec
r_1})} {\vec v}_{\perp} \frac{(e^{i\omega \delta t (1-{\hat
n}\cdot{\vec \beta}n)}-1)}{1-{\hat n}\cdot{\vec \beta}n}
\label{eq:pulse2} \ee
where $\mu_r$ is the relative permeability and $k=n\omega/c$.

The particle velocity ${\vec v}$ in the medium can be greater than
that of light.  The apparent singularity in Eq. (\ref{eq:pulse2}) at
$1-{\hat n}\cdot{\vec \beta}n=0$ defines the Cherenkov angle
$\theta_c$ as $\cos\theta_c\equiv 1/n\beta$.  However there is no
singularity, as seen by expanding the exponent.  Close to the
Cherenkov angle, Eq.  (\ref{eq:pulse2}) reduces to the form
\be R{\vec E}_{\omega}({\vec x})=\frac{i\omega}{\sqrt{2 \pi}}
\left(\frac{\mu_r q}{c^2}\right) e^{ikR}
e^{i\omega(t_1-\frac{n}{c}{\hat n}\cdot{\vec r_1})} {\vec v}_{\perp}
\delta t . \label{eq:pulse3} \ee 
Eqs. (\ref{eq:pulse2} \& \ref{eq:pulse3}), used by ZHS without
detailed derivation \cite{zhs92}, are coded into the simulation to
produce the field values for the tracks.  Eq. (\ref{eq:pulse3}) is
explicitly {\it linear} in the track length $|\vec v|\delta t$, a
feature we mentioned earlier.

Eqs. (\ref{eq:pulse2} \& \ref{eq:pulse3}) are incorporated into a
track-by-track Monte Carlo simulation.  The code uses the exact formula
Eq. (\ref{eq:pulse2}) unless the conditions are very close to the
Cherenkov angle, posing a $0/0$ numerical problem, in which case
Eq. (\ref{eq:pulse3}) is used.

Numerical summation of the electric field from all the tracks weighted
by the proper charge automatically incorporates the features of
coherence.  As illustrated shortly, coherence produces a signal that
peaks at the Cherenkov angle ($\theta_c$), with a width away from the
Cherenkov angle that shrinks with increasing frequency \cite{zhs92}.
The coherence features are sufficiently intricate that we have devoted
separate subsections to the topic.

\subsection{\it Parametrization Method}

Charges of opposite sign which radiate coherently will give electric
fields that cancel. It is the {\it excess} charge ($e-p$) that determines
the net field.  Following Buniy and Ralston (BR) \cite{roman00}, one
can then parameterize the {\it excess} (net) charge development in a
shower and calculate the electric field due to the prescribed
relativistic current.  This is a flexible and compact approach that
takes a few essential parameters from the shower and allows inspection
of both the Fresnel and Fraunhoffer zones.  We outline this method to
keep our presentation self-contained.

\begin{figure}
\vskip 5.5cm
\center \begin{picture}(0,0)
\includegraphics{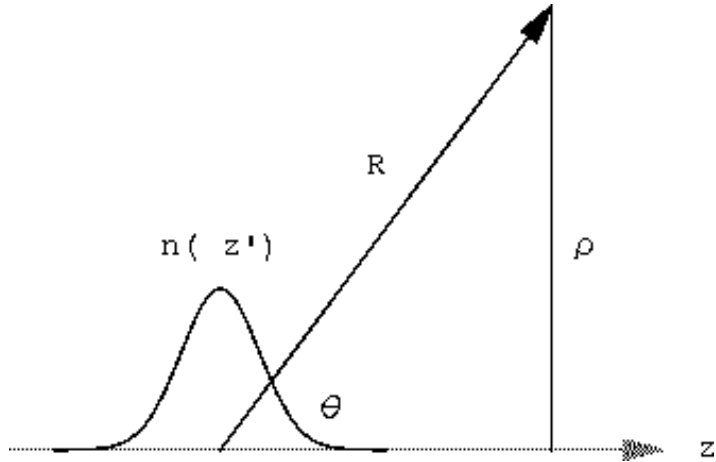}
\end{picture}
\vskip 0.5cm
\caption{\label{fig:diag2}
     \small 
Shower parametrization is done by assuming the longitudinal excess
charge development to be a Gaussian $n(z')$ moving along the shower
axis ($z$-direction) with speed $v$.  ${\hat n}$ is the direction of
the observer, making an angle $\theta$ with the shower axis.  $\rho$ is
the transverse distance to the observer.}
\end{figure}

The vector potential ${\vec A}$ in the Lorentz gauge (adapted to
$\epsilon (\omega)$ and $\mu =1$) is given by
\be {\vec A}_{\omega}({\vec x}) = \frac{4 \pi}{c \sqrt{2 \pi}}
\int{\dd^3 x' \frac{e^{ik|{\vec x}-{\vec x'}|}}{|{\vec x}-{\vec
x'}|}} \int{\dd t' e^{i\omega t'} {\vec J}(t' {\vec x'})}
\label{eq:veca1} \ee
The BR method parametrizes the excess charge development in the
shower by a current along the shower axis ($z$) as
\be {\vec J}(t', {\vec x}')={\vec v}\;n(z') f(z'-vt', {\vec \rho}'), 
\label{eq:ral1} \ee
where $\rho=\sqrt{x^2+y^2}$ is the radial distance from the shower
axis, and $n(z')$ is the excess charge ($e-p$) distribution,
approximated by a Gaussian near the shower maximum (see
Fig. \ref{fig:diag2}), as
\be n(z)=\frac{n_{\rm max}}{\sqrt{2 \pi}}\,e^{-z^2/2 a^2}. 
\label{eq:ral2} \ee

This is analogous to Greisen and Rossi's parameterization
\cite{greisen41}, where $n_{\rm max}$ is the excess charge at the shower
maximum, except that the excess (not total) charge ($e-p$) is modeled 
in the BR case.  The parameter $a$ corresponds to the ``longitudinal 
spread'' of the shower near maximum and can be extracted by fitting a 
Gaussian to simulations of the excess charge profile of a shower.

The region of the shower over which the fields arrive nearly in phase
at a distance $R(t)$ is called the {\it coherence zone} \cite{roman00},
and is given by
\be \Delta z_{\rm{coh}} = \sqrt{\frac{R}{k \sin^2 \theta}},
\label{eq:ral3} \ee
where $k=2 \pi/\lambda$ is the wave number and $\theta$ is the angle
between ${\hat R}$ and the shower axis ($z$).  The value of the
dimensionless ratio $\eta$,
\be \eta=\left(\frac{a}{\Delta z_{\rm{coh}}}\right)^2 = 
\frac{k a^2}{R}\sin^2 \theta , \label{eq:ral4} \ee
determines how one should calculate the field.  The ranges $\eta < 1$
and $\eta > 1$ correspond to the Fraunhoffer and Fresnel zones,
respectively.  In the case of $\eta \geq 1$, the series expansion of
the phase, Eq. ($\ref{frapprox}$) fails, and the exact phase must be
kept.  As noted earlier, both the Fraunhoffer and Fresnel zones are
far-field problems in the sense that $kR \gg 1$ is assumed in both
cases.

For $\omega \sim$ GHz the Fraunhoffer approximation\footnote{Shown in
detail earlier (Eqs. (\ref{eq:pulse2} \& \ref{eq:pulse3})).} is
appropriate for $R>$ 100 m distances typical of the RICE experiment.
However, for a test beam experiment, where the interplay between
different length scales is important, the Buniy-Ralston method to
calculate the field in the Fresnel zone is required.

\subsection{\it Radiated Power and The Form Factor}

We return to the ansatz Eq. (\ref{eq:ral1}), which describes a current
with transverse distribution independent of the longitudinal
evolution.  This is a reasonable assumption after the first few
radiation lengths of the shower development.  We make the specific
ansatz
\be {\vec J}(t', {\vec x'}) = {\hat e}_z q\,v\,n(z')\,g({\vec \rho'})
\,\delta (z'-vt'). \label{eq:simpcurrent} \ee
Note that the transverse extent of showers is small compared to the
longitudinal extent.  The phase and the denominator in Eq.
(\ref{eq:veca1}) can be safely expanded in cylindrical coordinates as
\ba |{\vec x}-{\vec x'}| &=& \sqrt{(z-z')^2 + \rho^2 - 2 {\vec \rho}
\cdot {\vec \rho'} + \rho'^2} \\
&\approx& R(z, z') \left(1 - \frac{\rho \rho'}{R^2} \cos
\phi' \right)+ {\cal O} \left( \frac{R'^2}{R^2} \right) \nn \ea
and
\be \frac{1}{|{\vec x}-{\vec x'}|} \approx \frac{1}{R} \ee
where $R(z, z')=\sqrt{(z-z')^2+\rho^2}$ and $R'=\sqrt{z'^2+\rho'^2}$.
This expansion of the transverse variables avoids the Fraunhoffer
approximation, and leads to a ``factored'' expression for the field:
\be R{\vec E}_{\omega}({\vec x}) \approx I \; F(\omega) . \ee
where $I$ is an integral characteristic of the longitudinal shower
history, and $F(\omega)$ is called the {\it form factor}.  The
integral $I$ can be evaluated in the saddle-point approximation, as
detailed elsewhere \cite{roman00}, or evaluated by Monte Carlo.  This
feature of {\it factorization} to express the field using a form
factor works in both the Fraunhoffer and Fresnel zones.

In the Fraunhoffer zone, there is a further, and very remarkable,
simplification.  Keeping the term linear in $z'$, and dropping terms of
order $R'^2/R^2$ in the phase expansion, the electric field at the
Cherenkov angle ($\theta_{\rm c}$) is
\ba \label{eq:field} 
R{\vec E}_{\omega}({\vec x}) &=& q \; 2\sqrt{2\pi}\;i e^{ikR}
\left[{\hat e}_z \left( \frac{\omega}{c^2}-\frac{k \cos\theta_{\rm
c}}{\epsilon v} \right) -{\hat e}_{\rho}\frac{k \sin\theta_{\rm
c}}{\epsilon v} \right] \\ && \times \int {\dd z'} n(z')
\int {\dd \rho'} \rho' g(\rho') \int {\dd \phi'}
e^{-i\frac{n \omega}{c}\rho'\sin \theta_{\rm c} \cos \phi'}
\nn \ea
It is remarkable that all terms in the phase exponent depending on
$z'$ have vanished.  This means that radiation from the shower is
coherent over the entire length of the shower in the Fraunhoffer
approximation: the ``coherence zone'' is limited only by the track
length\footnote{The importance of the coherence zone is well known in
the Landau-Pomeranchuk-Migdal \cite{landau53-1, landau53-2, migdal56}
effect, where the lengthening coherence zone suppresses
bremsstrahlung.  The Cherenkov condition causes the coherence zone to
expand to equal the entire track length.}.  In obtaining these
results, the Cherenkov condition $\cos \theta_{\rm c} = c/nv =
1/n\beta$ and the approximations: $\cos \theta_{\rm c}=z/R$, $\sin
\theta_{\rm c} = \rho/R$ (see Fig.  \ref{fig:diag2}) and $\rho'/R \ll
1$ were used.

Inspecting the transverse integral, we see that it {\it alone}
contributes to the form factor $F(\omega)$, which is given by
\ba \label{eq:form2} F(\omega)&=&\int \dd^{2} x_{\perp}' \, e^{-i \vec
k_{\perp} \cdot \vec x_{\perp}' } g(\vec x_{\perp}'), \nn \\*
&=&\int^{\infty}_0 {\rm d\rho'} \rho' g(\rho') \int^{2\pi}_0 {\rm
d\phi'} \; e^{-i\frac{n\omega}{c} \rho'\sin \theta_{\rm c}
\cos \phi'} ,\\ 
&=&  2 \pi \int^{\infty}_0 {\rm d\rho'} \rho' g(\rho') \; J_{\rm o}
\left( \frac{n\omega}{c} \rho' \sin \theta_{\rm c} \right). \nn \ea
Here $J_{\rm o}$ is the Bessel function of order unity.  Just as in
particle physics usage, $F(\omega)$ is the Fourier transform of the
(transverse) excess charge distribution.

The preceding analysis leads to a convenient expression for the
electric field at the Cherenkov angle in the Fraunhoffer
approximation, which apart from some normalization factors is given by
\be |R{\vec E}_{\omega}({\vec x})| = q\;2\sqrt{2\pi}\,I(a)\,\frac{\omega
\sin \theta_{\rm c}}{c^2}\,|F(\omega)|. \label{eq:amplitud} \ee
The radiated power at the Cherenkov angle ($\theta_{\rm c}$) depends
on the form factor only.  As noted earlier, the same form factor can
be used in both the Fraunhoffer and Fresnel limits, so that laboratory
information about the form factor can be used directly.  The form
factor is the central topic of subsequent sections discussing the
coherence.

\section{Electric Field Pulse Calculation}

In this section we carry out the computational outline just developed.
The vector nature of superposition is taken into account.  The
analytic construction of the form factor, developed earlier, is
adapted to the numerical calculation, so that two independent methods
can be compared.  We observe using the simulation data that coherence
extends far above the frequency regime anticipated from simple
estimates using characteristic scales such as the Moliere radius.

\subsection{\it Vector Superposition}

To calculate the electric pulse from a GEANT generated shower, we
summed the contributions to the electric field in Eqs.
(\ref{eq:pulse2} \& \ref{eq:pulse3}) from all charged track segments
using full 3-dimensional geometry.  We denote the observation point by
a unit vector, ${\hat n}=(\sin\theta \cos\phi,\; \sin\theta
\sin\phi,\; \cos\theta)$.  The starting time $t_1$ of each track
segment is obtained from GEANT output.  The time interval $\delta t$
for each track-segment is calculated for a particle of total energy
$E$ (from GEANT output) traveling at constant speed $\beta$ throughout
the track segment.  Slowing of low energy particles is thus taken care
of approximately.  The electrical field has an azimuthal symmetry
which will be investigated later.

The electric field is a complex vector quantity.  As such there are
two questions of coherence, one to do with phase and one to do with
the vector nature of the field.  The vector nature of the field
depends on the term ${\hat n}\times({\hat n}\times{\vec v})=-{\vec
v}_{\perp}$ in Eqs.  (\ref{eq:pulse2} \& \ref{eq:pulse3}).  This
construction picks out the component of the velocity which is
perpendicular to the direction of the observer.  Since the track
segments vary in direction, we add the electric field contributions as
vectors which allow any cancellation that may occur.  The electric
field amplitude is then proportional to the track length transverse to
the direction to the observer (${|\vec v|}_{\perp} \delta t$).  On the
Cherenkov cone, one then expects that the field generated by the
entire shower is proportional to the track length projected along the
shower axis times the sine of the Cherenkov angle.  The projected
track length thus accounts for the vector nature of the electric field
for tracks that go in different directions, and serves as an important
diagnostic of the whole procedure.

However, the projected track length says nothing in itself about the
conditions for phase coherence, which is the topic of the next
subsection.

\subsection{\it Phase Coherence}

The numerically generated phases of the fields, track-by-track, are
determined by the complex exponentials in Eq.  (\ref{eq:pulse2}).  We
call the phase angles $\omega (t_1 - \frac{n}{c} {\hat n} \cdot {\vec
r_1})$ and $\omega \delta t (1- {\hat n} \cdot {\vec \beta} n)$ the
{\it translational phase} (TP) and the {\it Cherenkov phase} (CP)
respectively.  The translational phase is kinematic, a consequence of
translational invariance, and depends on the beginning of each track
segment.  The Cherenkov phase vanishes at the Cherenkov angle, which a
point of stationary phase and dominates the emission.  Two or more
track segments will contribute in phase if the observer lies on the
Cherenkov cone, and the beginnings of the tracks do not destructively
interfere.

\begin{figure}
\vskip 7.5cm
\center
\begin{picture}(0,0)
\includegraphics{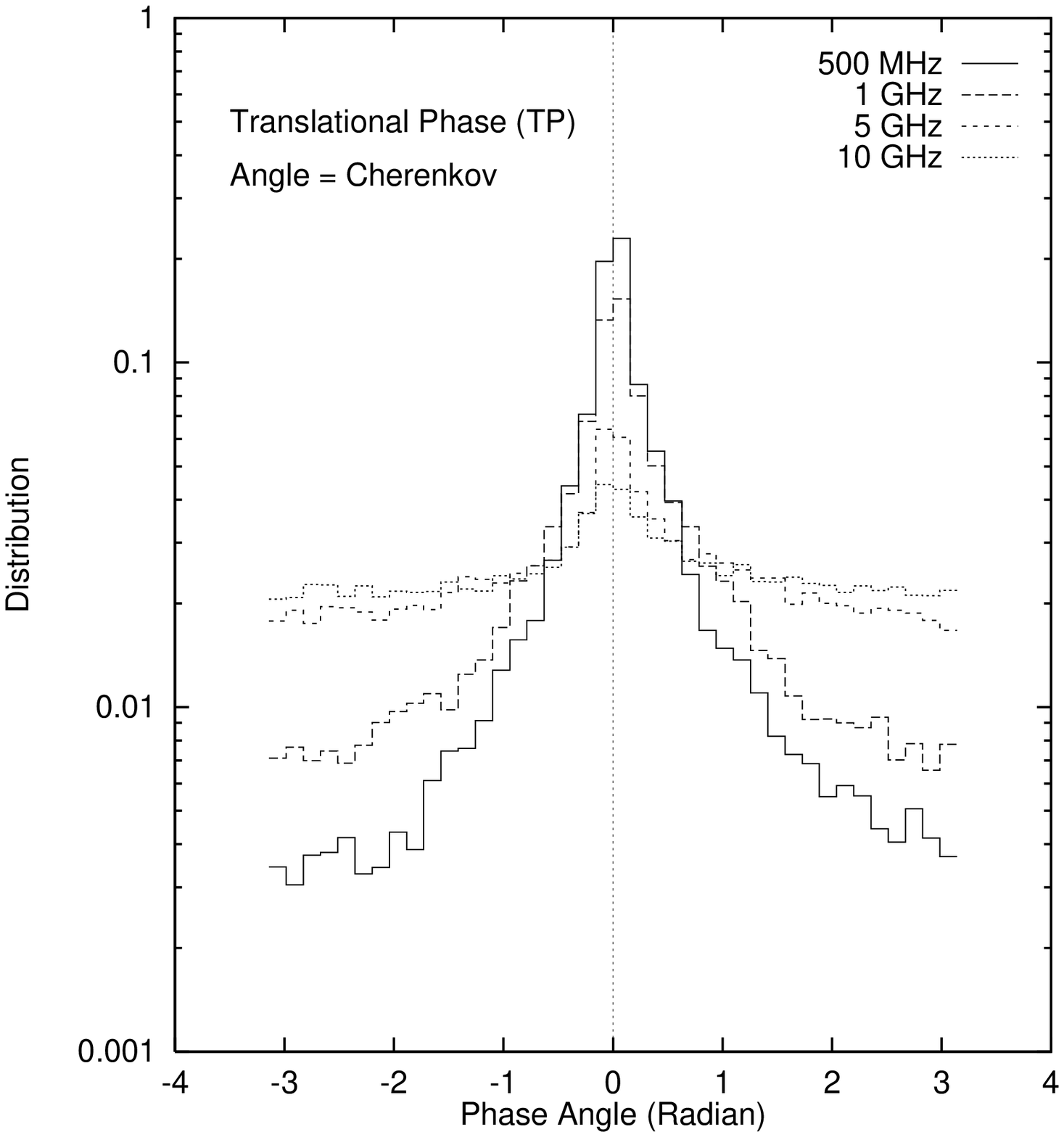}
\end{picture}
\begin{picture}(0,0)
\includegraphics{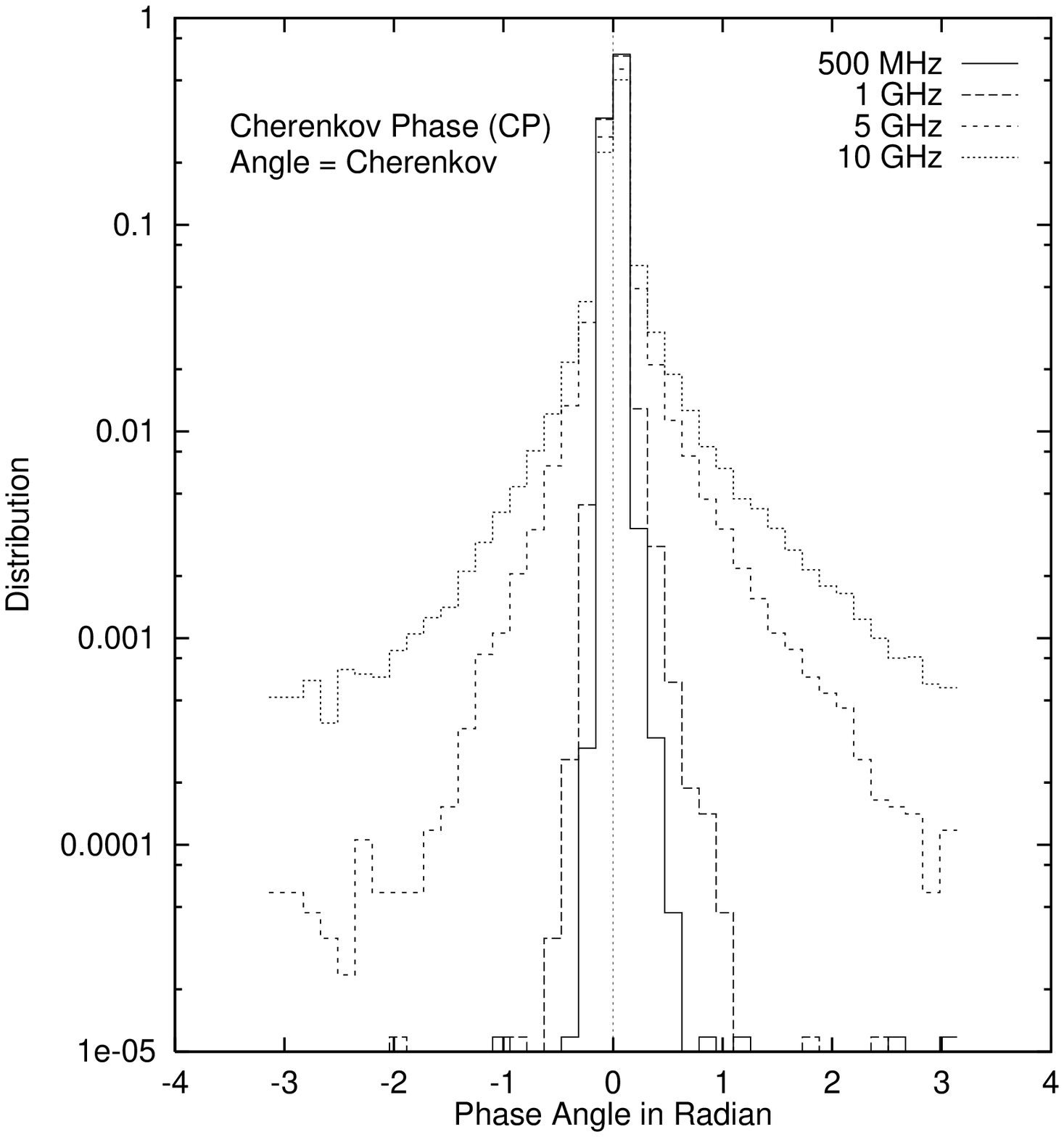}
\end{picture}
\vskip -0.5cm
\caption{\label{fig:phase-c}
\small Distributions of the translational phase (TP) and the Cherenkov
phase (CP) at the Cherenkov angle ($\theta = \theta_{\rm c}$) for a
single shower of energy 100 GeV and threshold of 0.611 MeV.  The TP
(left plot) shows strong coherence (sharp peak) at low frequencies.  A
slight positive enhancement of the TP can be interpreted as particle
positions slightly lagging the light cone because $\beta<1.$ The CP
(right plot), in comparison, remains strongly peaked at all
frequencies, which underscores the argument that the phase (CP) varies
a little over the whole track.}
\end{figure}

\begin{figure}
\vskip 7.5cm
\center
\begin{picture}(0,0)
\includegraphics{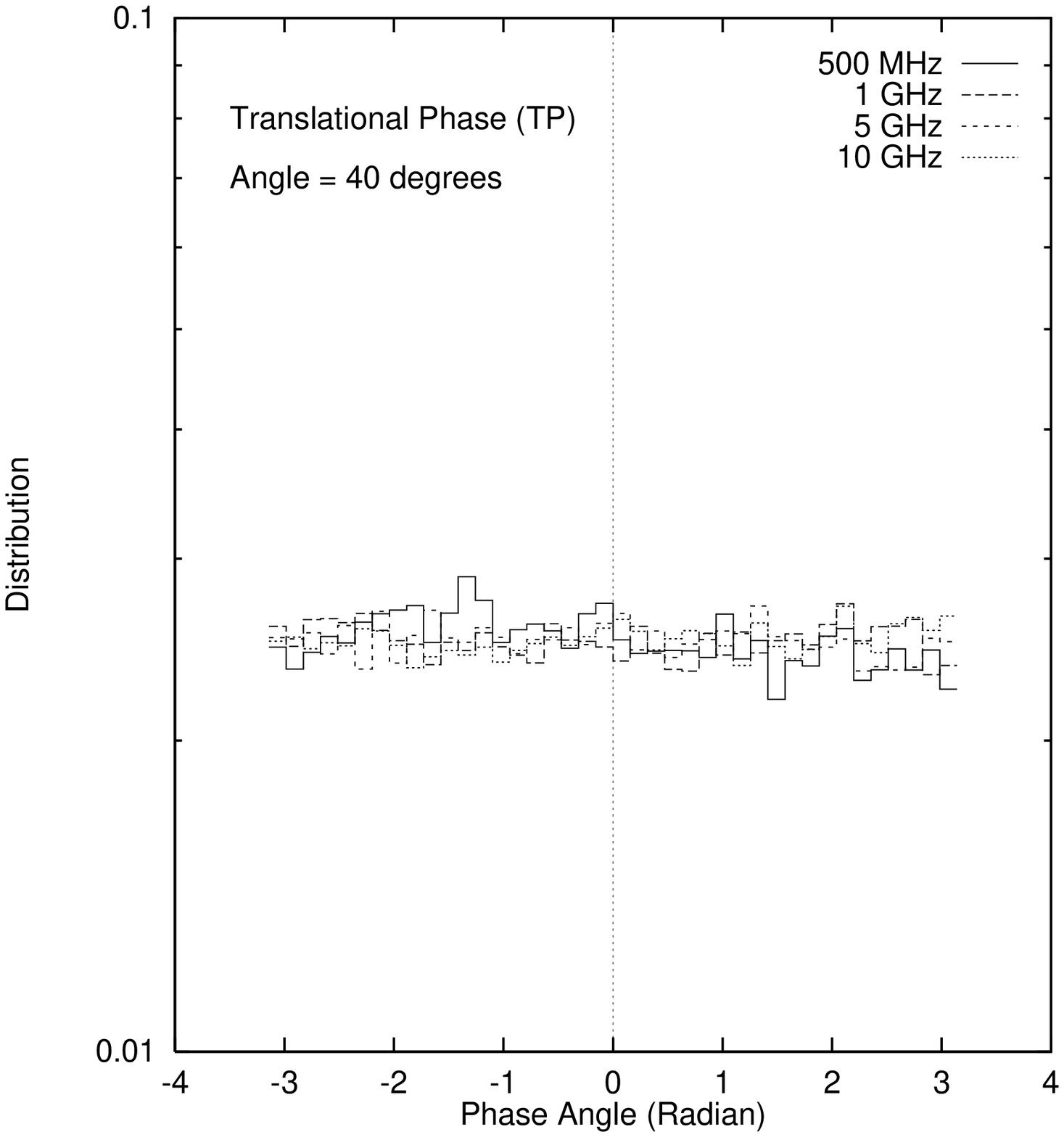}
\end{picture}
\begin{picture}(0,0)
\includegraphics{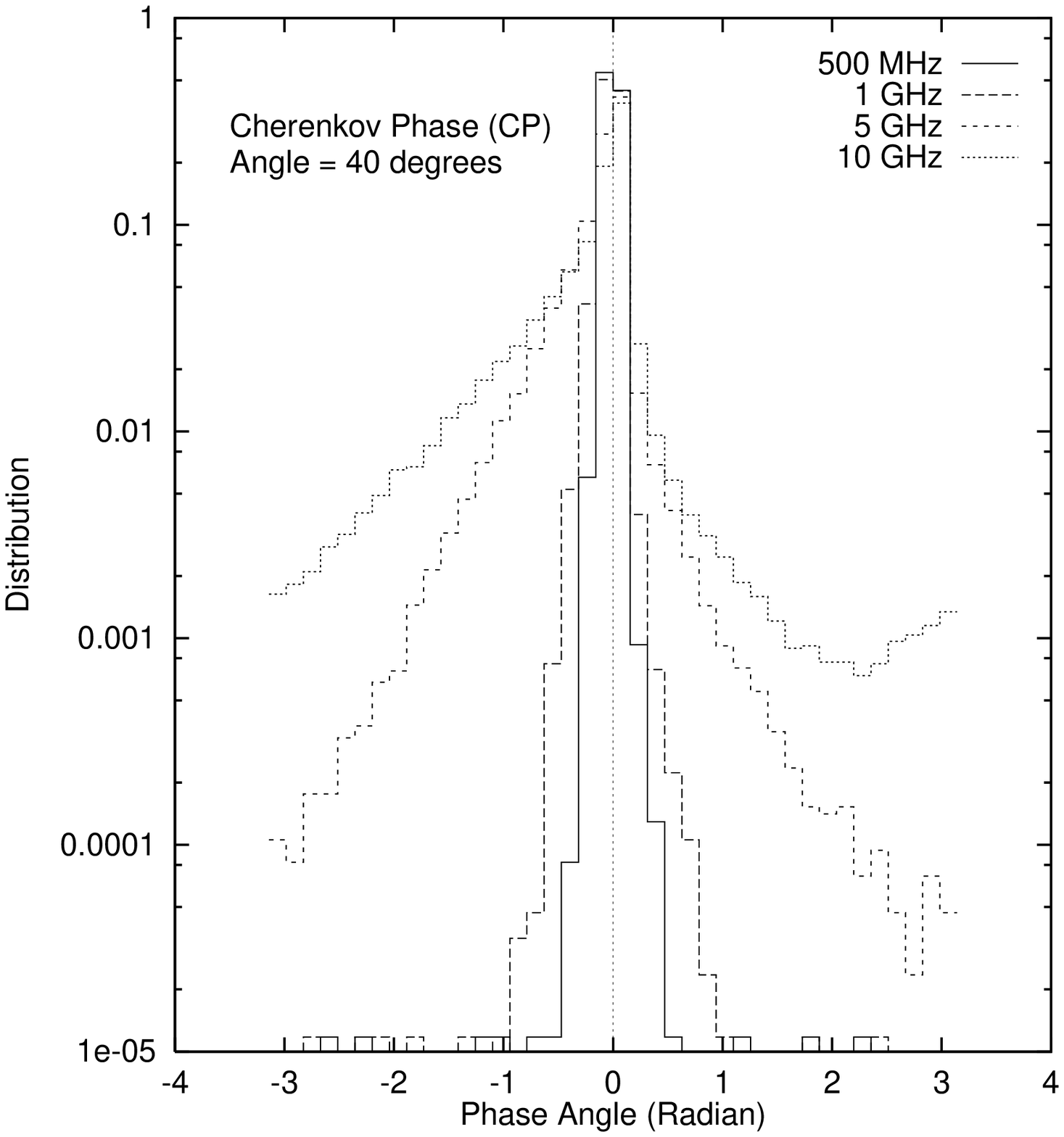}
\end{picture}
\vskip -0.5cm
\caption{\label{fig:phase-40}
\small Distributions of the TP and the CP at an angle ($\theta =
40^{\circ}$) off the Cherenkov cone for a single shower of energy 100
GeV with total energy threshold of 0.611 MeV. The flat distributions
of the TP (left plot) at all frequencies indicates the randomness of
phases coming from randomly located track segments.  The CP
distribution (right plot), is qualitatively the same as the case on
Cherenkov angle.  Taken together, these features indicate that phase
coherence is dominated by the TP.}
\end{figure}

We studied the distribution of the TP and CP at the Cherenkov angle
($\theta \approx 55.8^{\circ}$) and at an angle ($\theta =
40^{\circ}$) off the Cherenkov angle ( Figs.  \ref{fig:phase-c} \&
\ref{fig:phase-40}).  We used a single 100 GeV shower with 0.611 MeV
total energy threshold to make the phase distribution plots.  We
studied the distribution at frequencies: 10 GHz, 5 GHz, 1 GHz and 500
MHz.

The distribution of the TP (Fig.  \ref{fig:phase-c}) shows a strong
peak at low frequency, indicating coherent phase emission from track
segments at the Cherenkov angle.  At high frequencies, the
distribution tends to become flat, with random phases indicating a
loss of coherence.  A net positive phase indicates the particle
positions slightly lagging behind the light cone due to their speed
$\beta$ slightly less than 1.  The flat distribution of the TP in Fig.
\ref{fig:phase-40}, on the other hand, clearly indicates random phases
coming from track segments at angles off the Cherenkov cone.

The distribution of the CP remains qualitatively the same both at the
Cherenkov angle (Fig. \ref{fig:phase-c}) and at angles off the
Cherenkov angle (Fig. \ref{fig:phase-40}).  It is also apparent that
frequency dependence of the CP is very weak.

\begin{figure}
\vskip 5.cm
\center
\begin{picture}(0,0)
\includegraphics{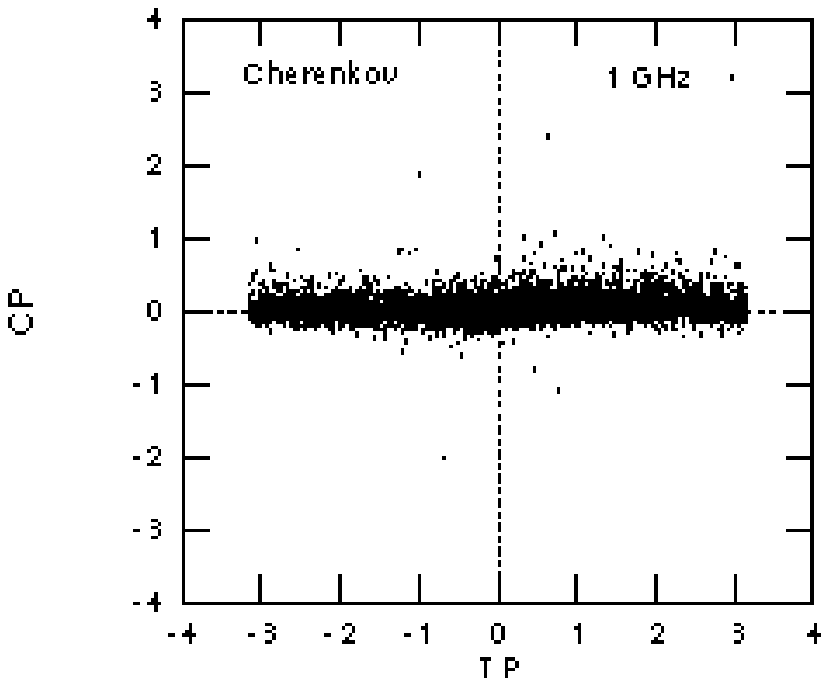}
\end{picture}
\begin{picture}(0,0)
\includegraphics{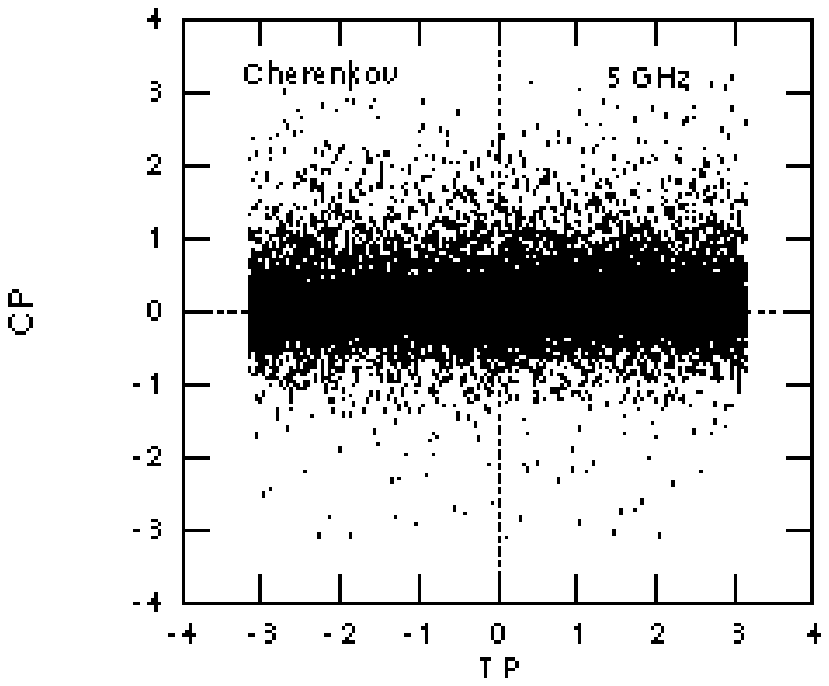}
\end{picture}
\vskip 0.5cm
\caption{\label{fig:scatter-c}
         \small
Scatter plots of the translational phase (TP) and the Cherenkov phase
(CP) at the Cherenkov angles ($\theta_{\rm c}$) and at frequencies: 1
GHz (left plot) and 5 GHz (right plot).  A correlation between the two
variables would appear as a slanted line, or similar feature.  The
plots demonstrate that the translational and Cherenkov phases are
substantially uncorrelated.  }
\end{figure}

\begin{figure}
\vskip 5.cm
\center
\begin{picture}(0,0)
\includegraphics{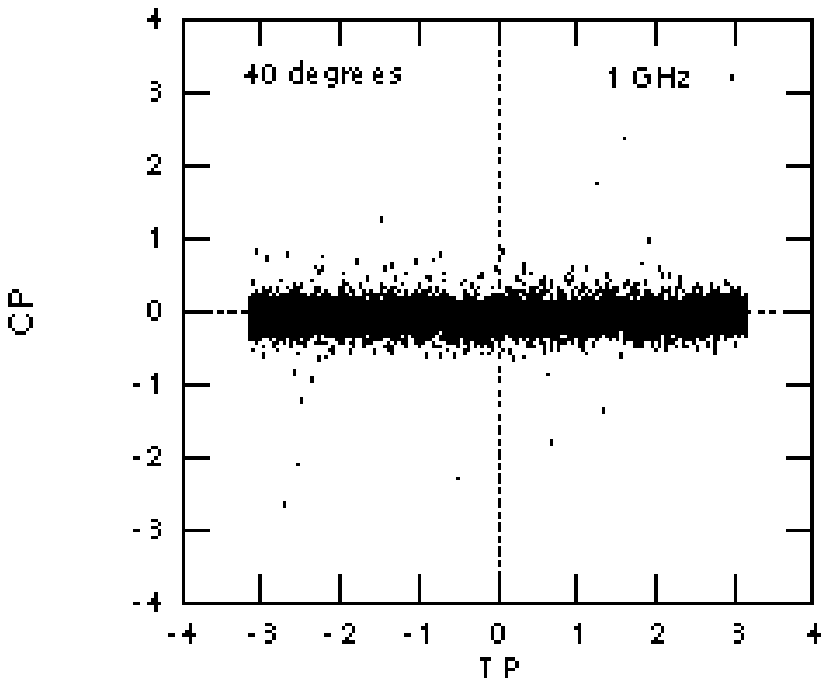}
\end{picture}
\begin{picture}(0,0)
\includegraphics{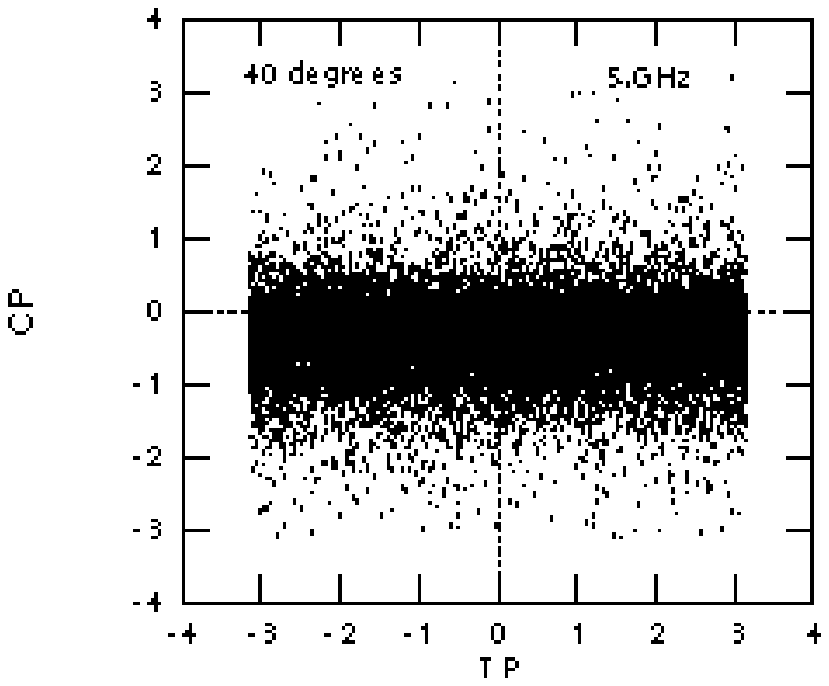}
\end{picture}
\vskip 0.5cm
\caption{\label{fig:scatter-40}
         \small
Same as Fig. \ref{fig:scatter-c} at angle ($\theta = 40^{\circ}$).  }
\end{figure}

The preceding discussion makes clear that {\it phase coherence is
dominated by the} TP.  Finally, the distribution of TP and CP
variables are substantially {\it uncorrelated} (see
Figs. \ref{fig:scatter-c} \& \ref{fig:scatter-40}).  These features
indicate a valid factorization of the electric field, leading to an
independent motivation for the Ansatz Eq. (\ref{eq:simpcurrent}) and
consequent appearance of the form factor.  We discuss this
factorization in the following subsection.

\subsection{\it Factorization of the Electric Field, and the
Discrete Form Factor}

The electric field from all the shower particles can be factorized
because the TP and CP phases are uncorrelated, as shown above.  We
write the field in factored form as
\be E_{\omega}^{\rm tot} \propto \sum_{j}e^{i \phi^{\rm TP}_{j}}( e^{i
\phi^{\rm CP}_{j}}-1) \sim \sum_{j} e^{i \phi^{\rm TP}_{j}} \sum_{j}( e^{i
\phi^{\rm CP}_{j}}-1) \;,
\label{eq:factorize}
\ee
where the sums are over all track segments.  The factorization takes a
rather simple form at the Cherenkov angle.  The electric field
close to the Cherenkov angle in Eq. (\ref{eq:pulse3}) can be written
approximately for a single track as
\be R{\vec E}_{\omega}({\vec x})=\frac{i\omega}{\sqrt{2\pi}}
\left(\frac{q}{c^2}\right) {\vec v}_{\perp} \delta t \; e^{i kR} \; 
e^{i\omega z_1(\frac{1}{v} - \frac{n}{c}\cos \theta)} \; e^{
-i\frac{n\omega}{c}\,{\hat n} \cdot {\vec r}_{1\perp} }.
\label{eq:factor-c1} \ee
The approximation $t_{1} \simeq z_{1}/v$ has been assumed in writing
Eq. (\ref{eq:factor-c1}), which is well supported by the TP phase plot
at the Cherenkov angle shown above.  The phase angle $\omega z_1(1/v
- n \cos \theta/c)$ vanishes for the Cherenkov condition
($\cos \theta_{\rm c} = c/nv$) and we have full coherence along the
$z$-direction (shower axis) as observed earlier.

We now write the factorization in Eq. (\ref{eq:factorize}) for the
total electric field at the Cherenkov angle from all the charged
particles in the shower as
\be R{\vec E}_{\omega}^{\rm tot}({\vec x}) =
\frac{i\omega}{\sqrt{2\pi}} \left(\frac{q}{c^2}\right)\;e^{i kR} 
\sum_j s^j ({\vec v}_{\perp})^j \delta t^j \; e^{
-i\frac{n\omega}{c}\,{\hat n} \cdot ({\vec r}_{1\perp})^j}
\label{eq:factor-c2} \ee
where $s^j = \pm 1$ for positrons and electrons respectively.  The
total electric field in Eq. (\ref{eq:factor-c2}) is thus proportional
to the total track length ($\sum_j s^j ({\vec v}_{\perp})^j \delta
t^j$) transverse to the direction of the observer at any frequency
$\omega$.  This track length is the projected ($e-p$) track length
times $\sin \theta_{\rm c}$, as described earlier.

The coherent electric field emission at different frequencies now
depends on the Monte Carlo ``discrete form factor'' $ F(\omega)_{\rm
MC}$, given by
\be F(\omega)_{\rm MC} = \sum_j s^j e^{ -i\frac{n\omega}{c}\,{\hat
n} \cdot ({\vec r}_{1\perp})^j} = \sum_j s^j e^{ -i\frac{n\omega}{c}
x^j_1 \sin \theta_{\rm c}} .
\label{eq:mcformfactor} \ee
Note that the observation point is in the $x-z$ plane.

\subsection{\it The Discrete Form Factor and the Frequency Spectrum}

We calculated the discrete form factor in Eq. (\ref{eq:mcformfactor})
for 1 TeV, 500 GeV and 100 GeV showers.  A factor of $\omega$ times
the form factor is proportional to the electric field amplitude.

\begin{figure}
\vskip 7.5cm
\center
\begin{picture}(0,0)
\includegraphics{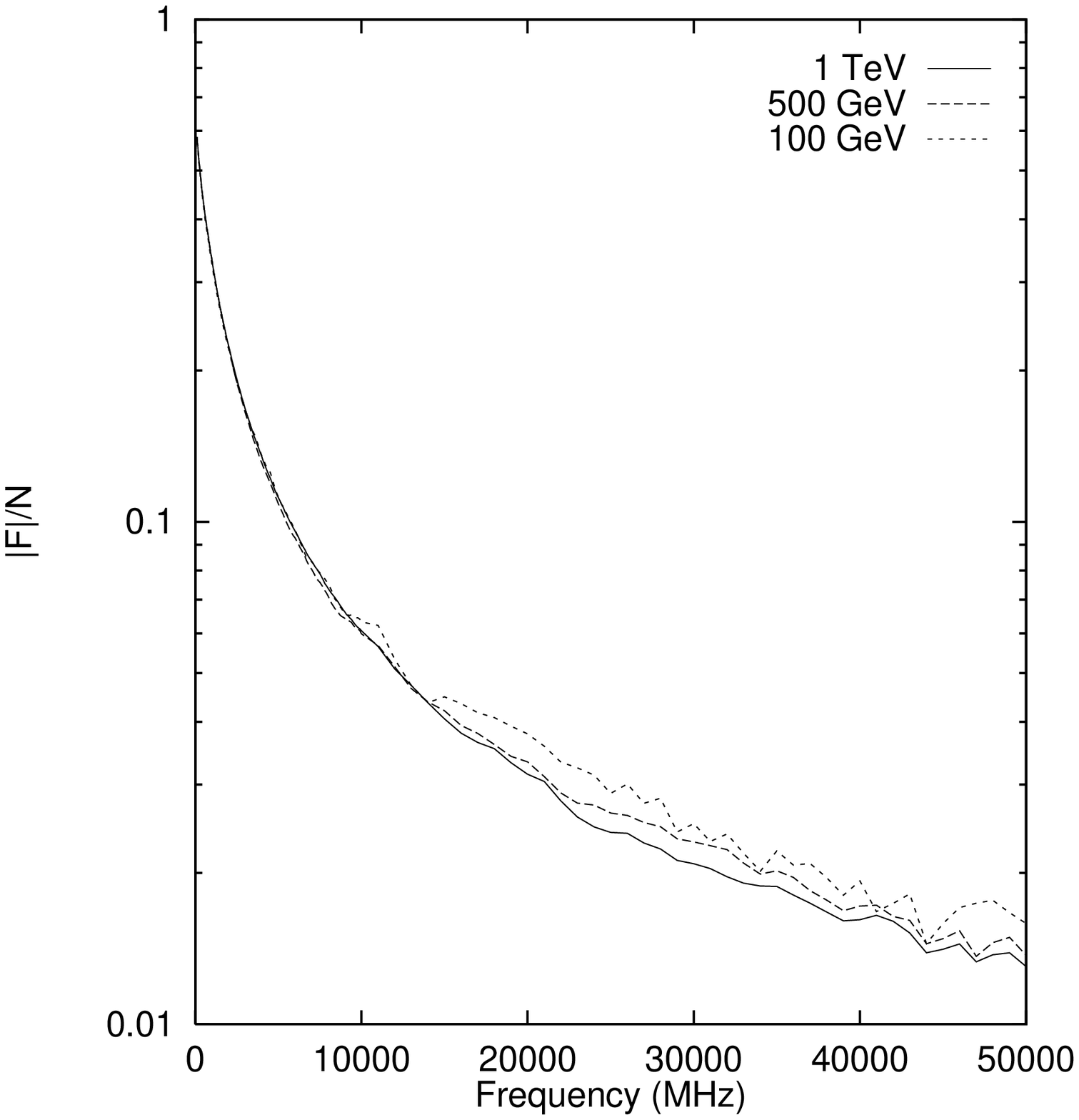}
\end{picture}
\begin{picture}(0,0)
\includegraphics{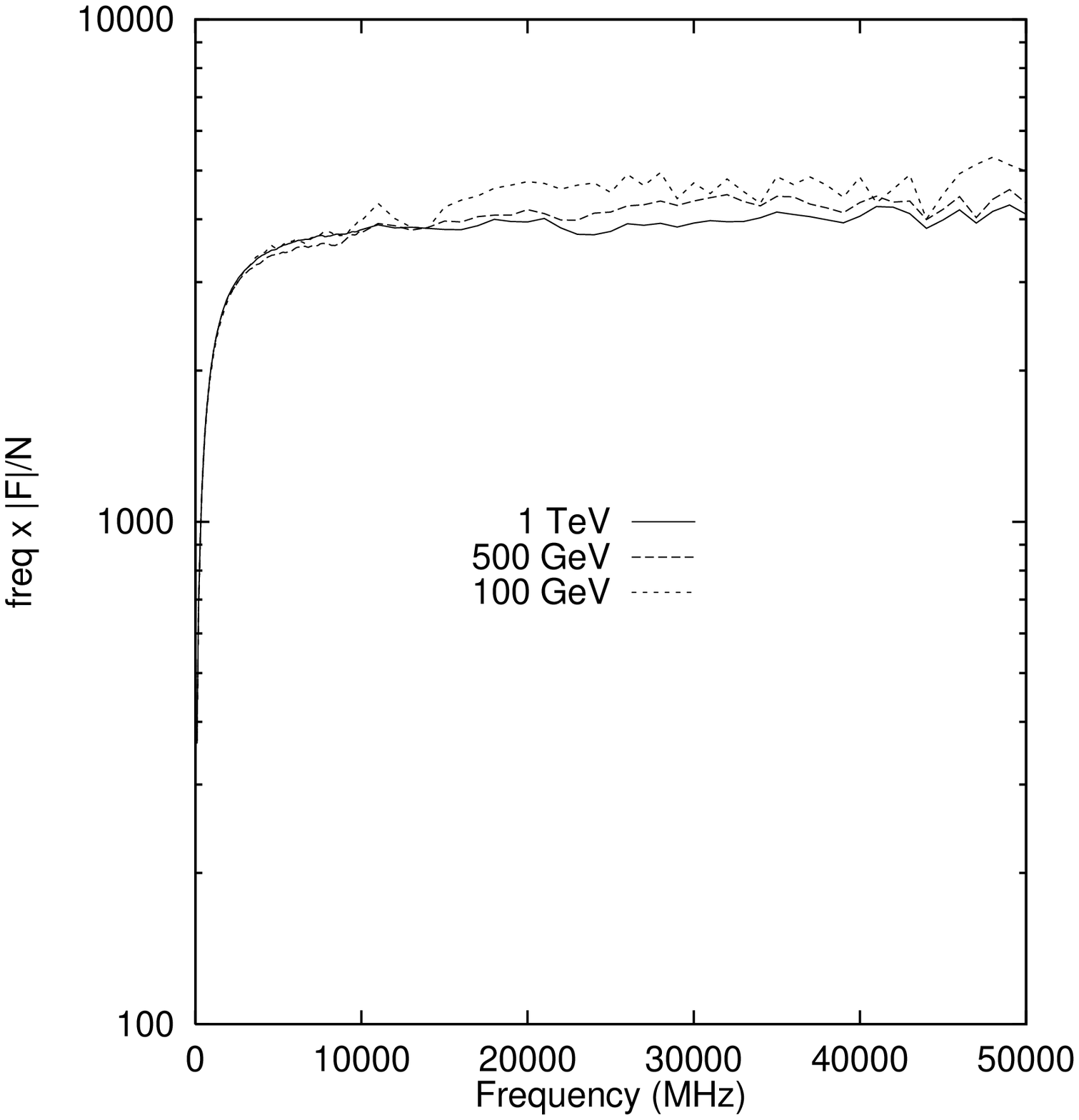}
\end{picture}
\vskip -0.5cm
\caption{\label{fig:form-all}
          \small
Absolute value of the form factor ($|F(\omega)|$) (left plot) and the
frequency spectrum ($\omega |F(\omega)|$) (right plot) plots for 1
TeV, 500 GeV and 100 GeV showers.  The form factor has been calculated
using Eq. (\ref{eq:mcformfactor}), where we naively sum over all the
shower particles along the $z$-axis (shower axis). The curves are
normalized by dividing by the particle count $N$.}
\end{figure}

First, we naively carried out the sum in Eq. (\ref{eq:mcformfactor})
for all the shower particles along the whole shower axis ($z$) as
suggested by the Cherenkov condition in the Eq.  (\ref{eq:factor-c1}).
The absolute value of the form factor and the frequency spectrum are
plotted in Fig.  \ref{fig:form-all}.  It shows an extended region of
coherence, with $|Form factor| \sim N/\omega$, where $N$ is the number
of shower particles.  This naive estimate is somewhat misleading
because the Cherenkov condition is not satisfied for the whole shower
axis.  The CP also becomes flat at very high frequencies, as does
the TP at the Cherenkov angle (see Fig. \ref{fig:phase-c}a).  A
further complication is statistical fluctuations, which become large
where the form factor is small.

For a better understanding of the form factor and the frequency
spectrum, we now turn to the analytic method.

\subsection{\it The Analytic Form Factor and the Frequency Spectrum}

The analytic form factor as defined in Eq. (\ref{eq:form2}) is the
Fourier transform of the snapshot of the charge distribution.  It is a
good approximation to assume that most of the electric field
contribution comes from the particles at the shower maximum
\cite{roman00}.

\begin{figure}
\vskip 6.75cm
\center
\begin{picture}(0,0)
\includegraphics{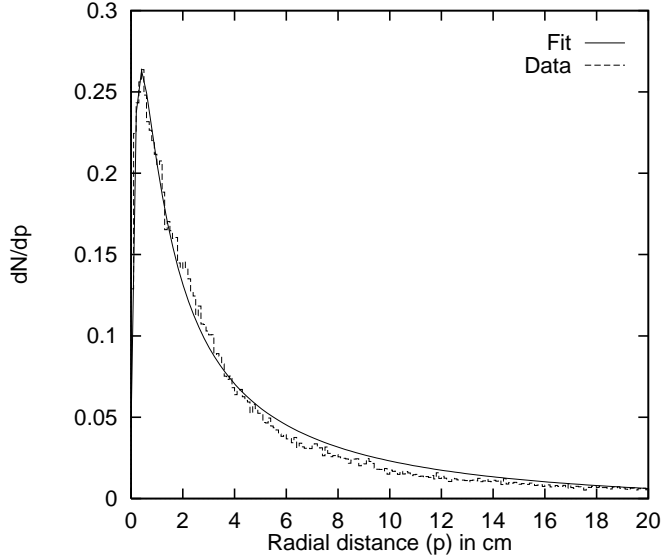}
\end{picture}
\vskip 0.5cm
\caption{\label{fig:padefit}
\small Radial distributions $\dd N/\dd \rho$ of the excess $(e-p)$ 
electrons within half a radiation length on both sides of the shower
maximum for a 100 GeV shower (averaged over 50 showers).  The solid
curve is a Pade(1,3) fit to the excess charge distribution.}
\end{figure}

To calculate the analytic form factor, we then determined the
transverse distribution of the particles within half a radiation
length on both sides of the shower maximum for a 100 GeV shower
(averaged over 50 showers).  We found both the distribution of the
total particles $\dd N(e+p)/\dd\rho$ and the distribution of the
excess electrons $\dd N(e-p)/\dd\rho$.  Here, $\rho$ is the
cylindrical radial distance from the shower axis.  The excess charge
distribution ($\dd N(e-p)/\dd\rho$) is plotted in
Fig. \ref{fig:padefit}.

We observed that the radial distribution falls approximately like
$1/\rho^2$ for large $\rho$, and peaks at a remarkably small value of
0.5 cm.  One way (by no means complete) to understand these results
recalls the multiple scattering of an electron by a static Coulomb
field.  For a screened Coulomb field, the modified Rutherford
scattering cross section is given by \cite{jackson}:
\be \frac{{\rm d} \sigma_s}{{\rm d} Q} = 8 \pi \left( \frac{Z
e^2}{\beta c} \right)^2 \frac{Q}{\left( Q^2+Q_s^2 \right)^2}
\label{eq:rutherford} \ee
where $Q$ is the momentum transfer and $Z$ is the atomic number of the
scatterer; $Q_s=(Z^{1/3}/192)mc$ is the momentum transfer associated
with the screening radius; $m$ is the mass of the electron.  For
elastic scattering in the ultra-relativistic limit, $Q^2 = 2
p^2(1-\cos\theta) \approx p^2 \theta^2$, where p is the electron
momentum.  The cross-section formula in Eq.  (\ref{eq:rutherford}) now
takes the form:
\be \frac{{\rm d} \sigma_s}{{\rm d} \theta} = 8 \pi \left( \frac{Z
e^2}{\beta c} \right)^2 \frac{p^2 \theta}{\left( p^2 \theta^2+Q_s^2
\right)^2} .
\label{eq:rutherford2}
\ee

The average deflection angle, $<\!\!\theta\!\!> \,\approx \rho/\Delta
z$.  We can get a crude estimate of the transverse distribution $\dd
N/\dd\rho$ from Eq. (\ref{eq:rutherford2}) as:
\be \frac{{\rm d}N}{{\rm d}\rho}= 8 \pi \left( \frac{Z e^2}{\beta c}
\right)^2 \left( \frac{n\Delta z^3}{p^2} \right) \frac{\rho}{\left(
\rho^2+\rho_{\rm o}^2 \right)^2}
\label{eq:rutherdist} \ee
where $n$ is the number density related to $N$ by the formula
$N=n\sigma \Delta z$.  The peak of this distribution is given by
\be \rho_{\rm o} = \frac{Q_s \Delta z}{p}. \ee

We evaluate this with the average energy of the particles taken to be
approximately equal to the critical energy ($E_{\rm c}$) at the shower
maximum.  The peak of the distribution is then $\rho_{\rm o}=0.5$ cm
for ice ($Z=7.2$, $\Delta z =$ 39 cm and $p \sim E_{\rm c}=$ 70 MeV).
The excellent agreement may be fortuitous, but gives some confidence
that the Monte Carlo excess charge distribution, which also includes
numerous atomic collision, pair production and Compton scattering
processes, has a simple physical origin.

We made a fit to the excess charge distribution with a Pade (1,3)
approximation of the form $\dd N/\dd\rho = f(\rho)=n (\rho - a) / (1 +
b\rho +c\rho^2 + d\rho^3)$ (see Fig. \ref{fig:padefit}).  The fitting
parameters are: $n=2.15$, $a=0.01$, $b=4.73$, $c=6.96$, $d=0.33$.  The
fit is good up to $\rho \sim 20$ cm.  The choice of the Pade fit was
made to preserve the $\rho^1$ geometric zero in $\dd N/\dd \rho$ at
the origin, and the $1/\rho^2$ asymptotic behavior.

Finally, we calculated the analytic form factor using
Eq. (\ref{eq:form2}) as:
\be F(\omega)=2\pi \int_0^{\infty} {\rm d}\rho \, f(\rho) \, J_{\rm o} 
\left(\frac{n\omega}{c} \, \rho \sin \theta_{\rm c} \right). \ee

\begin{figure}
\vskip 7.5cm
\center
\begin{picture}(0,0)
\includegraphics{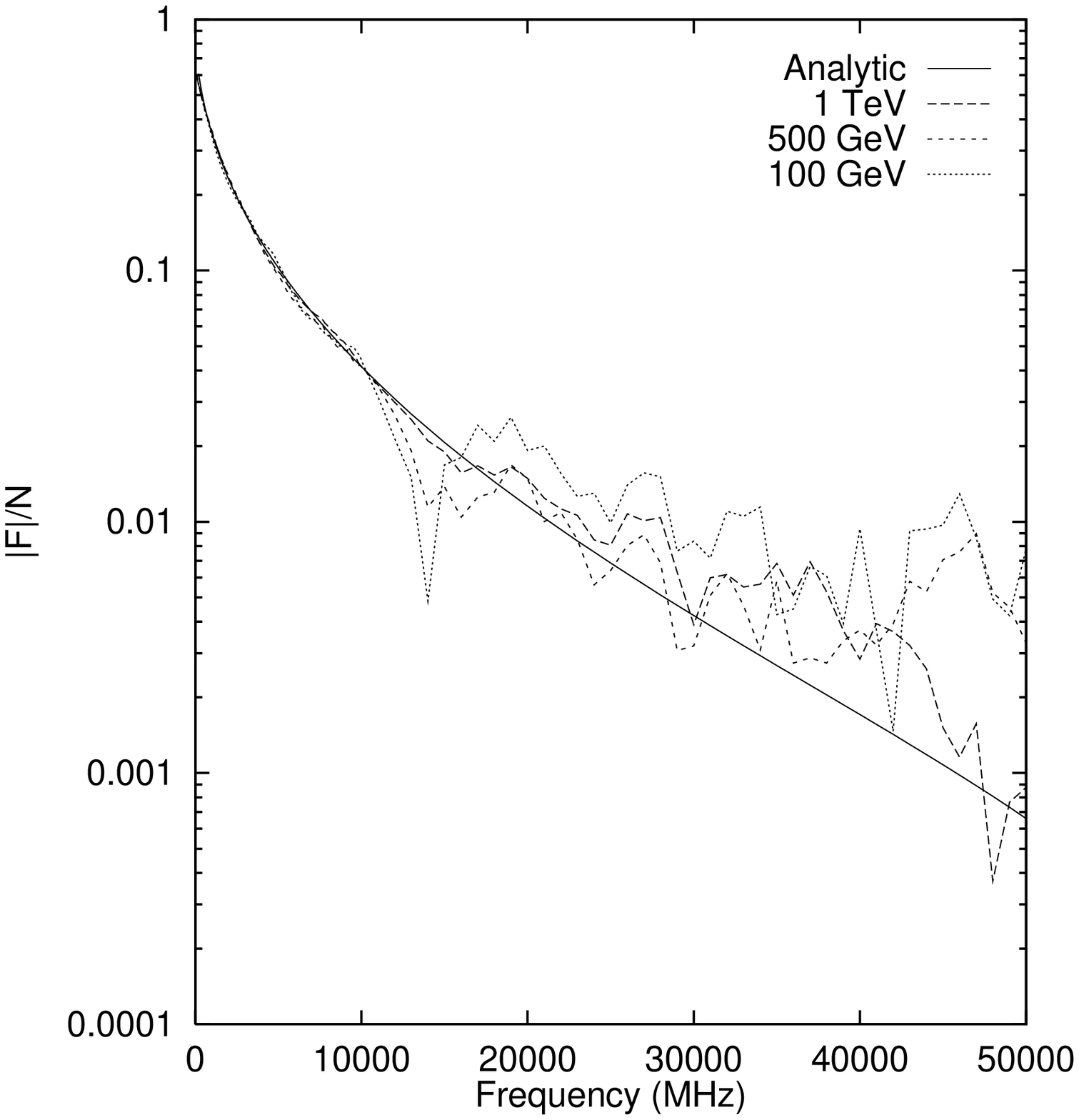}
\end{picture}
\begin{picture}(0,0)
\includegraphics{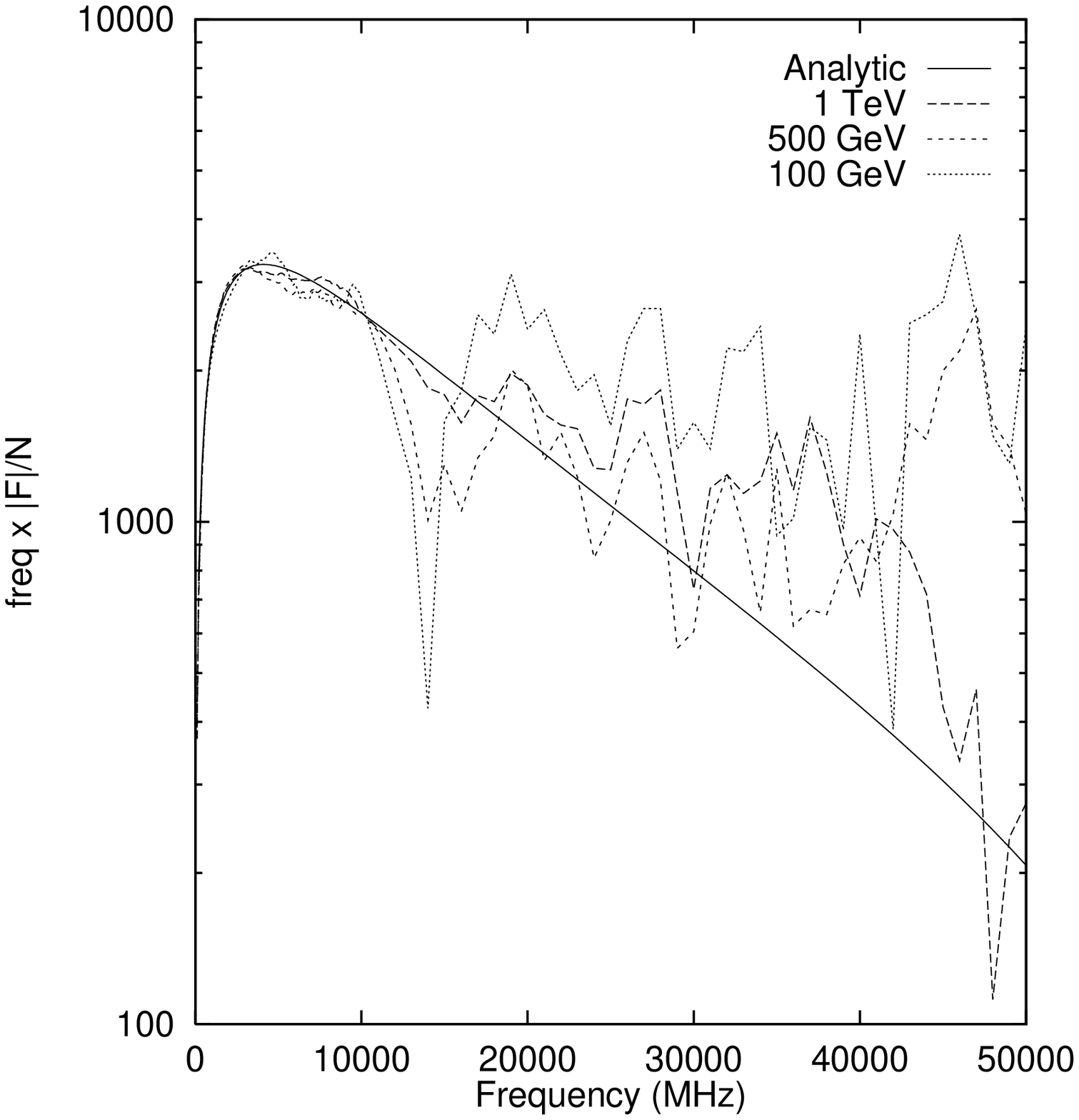}
\end{picture}
\vskip -0.5cm
\caption{\label{fig:form-max}
\small Absolute value of the form factor $|F(\omega)|$ (left) and the
frequency spectrum $\omega |F(\omega)|$ (right) plotted versus
frequency for 1 TeV, 500 GeV and 100 GeV showers.  The discrete form
factor has been calculated using Eq.  (\ref{eq:mcformfactor}), where
we sum over all the particles within a radiation length at the shower
maximum.  The analytic form factor has been calculated from the
Fourier transform to a fit to the excess charge distribution (see
Fig. \ref{fig:padefit}).  Both curves are normalized by dividing by
the particle count $N$ Note that the analytic curve tends to be lower
than the numerical calculation at high frequency.}
\end{figure}

For the high frequency end it was necessary to use a convergence
procedure to modulate the Bessel transform.  We compare the analytic
prediction to the discrete form factor of Eq.  (\ref{eq:mcformfactor})
at the shower maximum for 100 GeV, 500 GeV and 1 TeV showers.  The
comparison is shown in Fig.  \ref{fig:form-max}.  For reference we
also plotted the analytic and the discrete frequency spectra, which
are $\omega$ times the absolute value of the form factor.  The
agreement between the two methods is good up to 10 GHz for all
energies and up to 50 GHz for 1 TeV.

\subsection{\it Direct Calculation: Monte Carlo Field Spectrum}

\begin{figure}
\vskip 7.5cm
\center
\begin{picture}(0,0)
\includegraphics{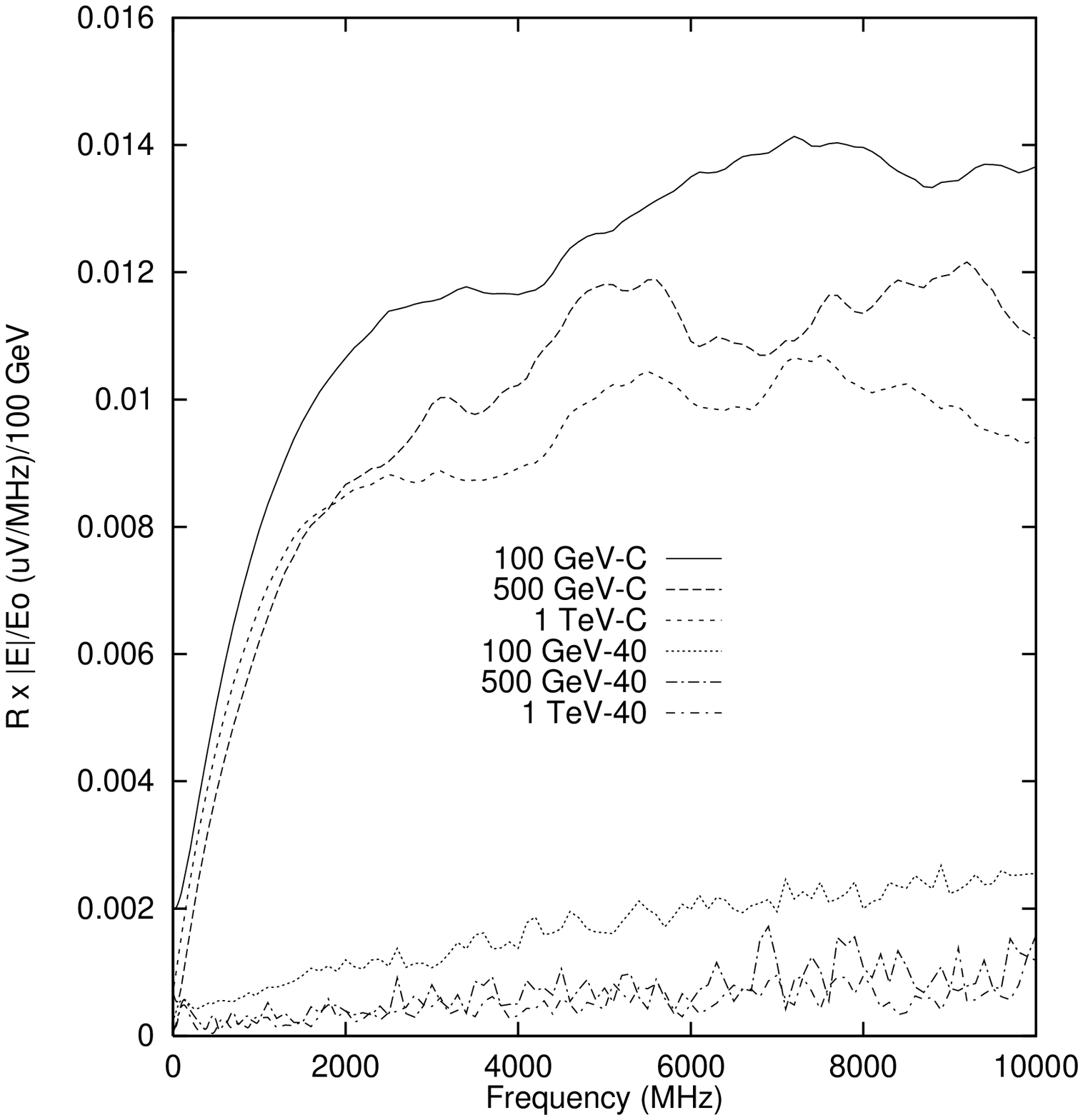}
\end{picture}
\begin{picture}(0,0)
\includegraphics{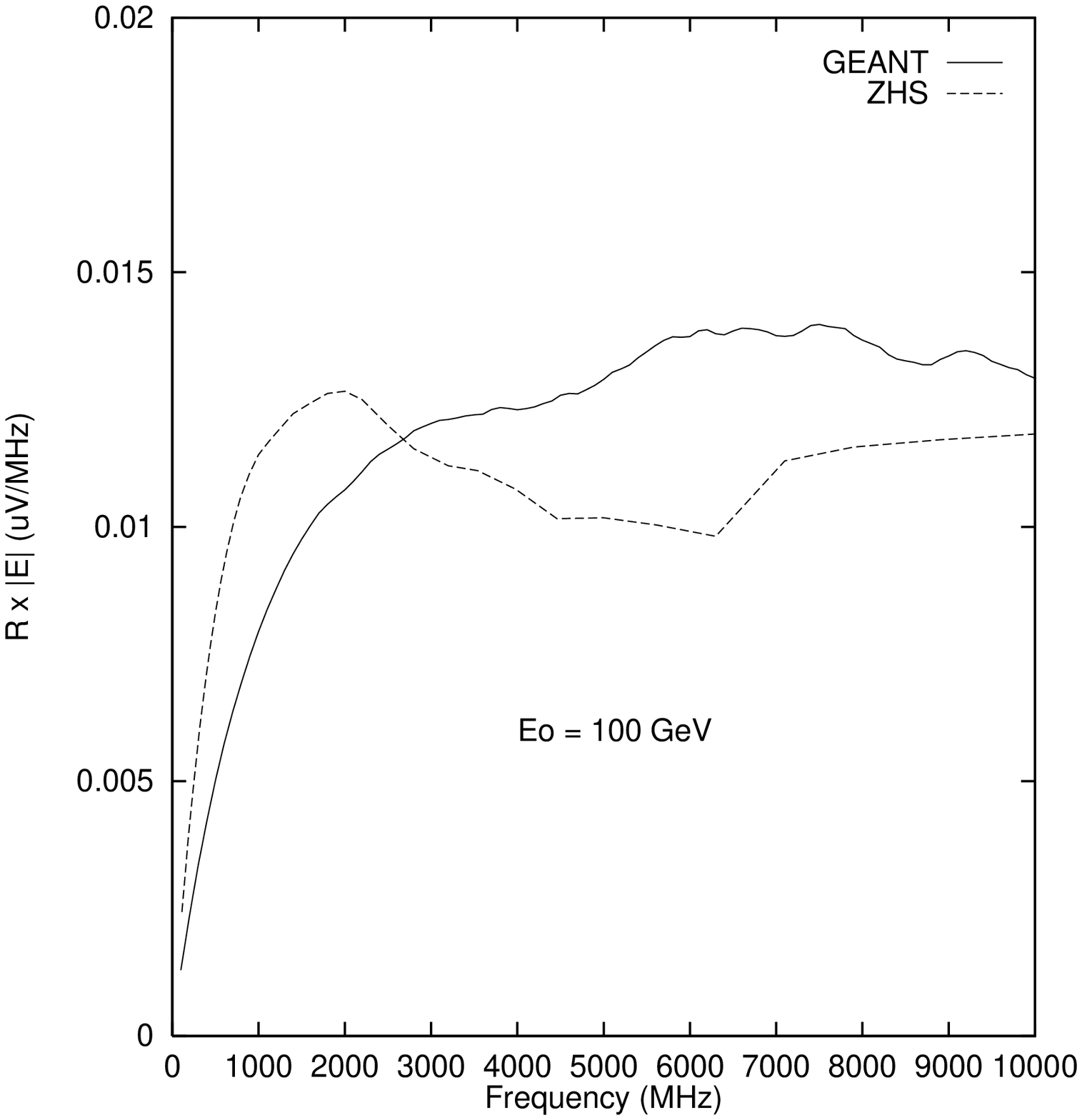}
\end{picture}
\vskip -0.5cm
\caption{\label{fig:freqspec1}
\small Direct Monte Carlo calculation of the frequency spectrum of the
electric field magnitude for 1 TeV, 500 GeV and 100 GeV showers
(left).  The calculation is done using Eqs. (\ref{eq:pulse2} \&
\ref{eq:pulse3}) at the Cherenkov angle and at $40^{\circ}$ angle.
The coherent behavior at the Cherenkov angle and the incoherent
behavior at the $40^{\circ}$ viewing angle are clear.  The frequency
spectrum at the Cherenkov angle for a 100 GeV shower is also compared
to the same calculated using ZHS code (right).  All thresholds are
0.611 MeV and the averages are done with 50 showers in case of 100
GeV, 20 showers in case of 500 GeV and 10 showers in case of 1 TeV.}
\end{figure}

The frequency spectrum of the electric field calculated using Eqs.
(\ref{eq:pulse2} \& \ref{eq:pulse3}) at the Cherenkov angle and an
angle off the Cherenkov angle are plotted in Fig.
\ref{fig:freqspec1}a.  We calculated the spectrum for 1 TeV, 500 GeV
and 100 GeV showers (each averaged over many showers).  The electric
field amplitude rises linearly at low frequency: this is the linear
dependence on $\omega$ due to dimensional analysis.  The coherent
behavior at the Cherenkov angle and the incoherent behavior at a
widely separated angle (40$^{\circ}$) are clear.  The frequency
spectrum of the electric field calculated from a 100 GeV GEANT
generated shower is compared to the same from the ZHS code in Fig.
\ref{fig:freqspec1}b.

\begin{figure}
\vskip 6.75cm
\center
\begin{picture}(0,0)
\includegraphics{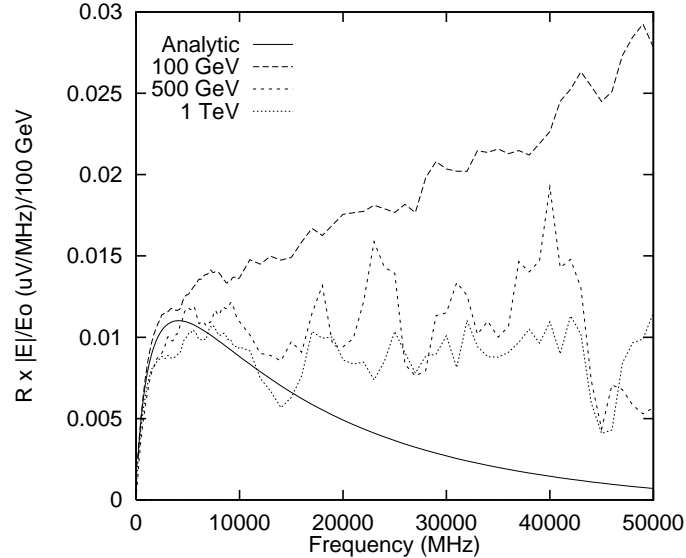}
\end{picture}
\vskip 0.5cm
\caption{\label{fig:freqspec2}
     \small 
Comparison between the frequency spectra of the electric field
magnitude at the Cherenkov angle from direct Monte Carlo calculation
and from analytic calculation.  The analytic calculation is done using
Eq. (\ref{eq:mcformfactor}) with the total projected track length of
70 m for a 100 GeV shower as found before.}
\end{figure}

From the previous analysis, the frequency dependence of the electric
field amplitude is expected to be $\omega |F(\omega)|$.  We compare
$\omega |F(\omega)|$ to $|E_\omega |$ generated by the Monte Carlo in
Fig.  \ref{fig:freqspec2}.  The electric field amplitude at the
Cherenkov angle is proportional to the {\it projected $(e-p)$ track
length} $\times \sin \theta_{\rm c}$ as shown in Eq.
(\ref{eq:factor-c2}).  We used the total projected ($e-p$) track
length value of 70 meter (as we found before) for a 100 GeV shower as
a pre-factor to normalize the analytic form factor.  It is clear from
the plot that the analytic form factor and simple dimensional
considerations explain the electric field amplitude rather well up
through 5-15 GHz range.  The coherence persists up to frequencies as
high as 50 GHz.  We notice that the agreement between the direct Monte
Carlo calculation of the frequency spectrum and the analytic frequency
spectrum gets better with increasing shower energy.  We believe that
the analysis captures the important physics of the processes, and
validates the results of the Monte Carlo.

\subsection{\it Related Issues}

Electrons and positrons undergo Coulomb scattering while traversing
through the medium.  The particle track is therefore not a straight
line from the point of its creation to the point where it falls below
threshold.  The track contains many {\it kinks} due to elastic Coulomb
scattering.  GEANT gives a detailed output of particle tracks which
contain these kinks as stated earlier; we calculate the electric field
from {\it step-tracks}.  The Monte Carlo developed by ZHS, on the
other hand used the straight tracks from the start points to the end
points, with timing correction, to calculate field \cite{zhs92} to
simplify their calculation.

\begin{figure}
\vskip 6.75cm
\center
\begin{picture}(0,0)
\includegraphics{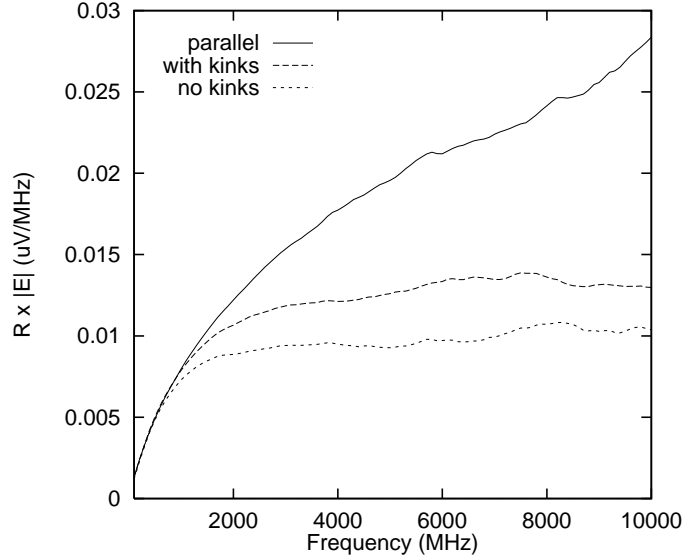}
\end{picture}
\vskip 0.5cm
\caption{\label{fig:spec-kinks}
     \small 
Frequency spectrum of the electric field magnitude at the Cherenkov
angle for a 100 GeV shower (averaged over 50 showers).  Our study
shows that the electric field calculated using particle tracks with
and without kinks make little difference.  The electric field
magnitude, on the other hand increases almost linearly with frequency
like a single charge when we took the component of the tracks parallel
to the shower axis only. }
\end{figure}

We studied the effect of these kinks and also the effect of taking all
tracks along the shower axis (see Fig. \ref{fig:spec-kinks}).  We
found little difference between the cases when we included (the usual
case) and not included the kinks.  This is a result of the extended
coherence zone along the shower axis at the Cherenkov angle.  When we
calculated the electric field taking only the components parallel to
the shower axis, we found that the field increases almost linearly
with frequency, as expected from a single charge.

For practical purposes, particles are removed from the Monte Carlo
simulation once they fall below certain threshold energy (0.611 MeV in
our case).  While physical particles do not suddenly stop moving,
simulated particles in the Monte Carlo, have potential emission of
stopping radiation.  Even though high energy particles are almost
parallel to the shower axis, the low energy particles close to
threshold are isotropic.  The stopping radiation (which is forward)
from all these isotropic particles in general could cause a problem if
they are aligned with the observer on the Cherenkov cone.

\begin{figure}
\vskip 6.cm
\center
\begin{picture}(0,0)
\includegraphics{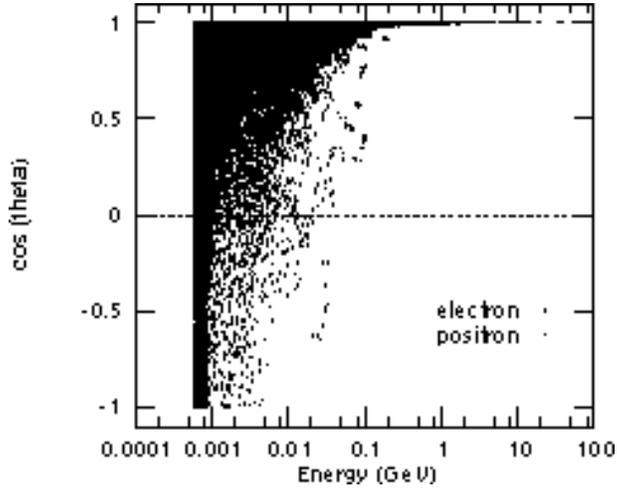}
\end{picture}
\vskip 0.5cm
\caption{\label{fig:polar-angle}
     \small 
Plot of the cosine of the polar angle ($\theta$) of each track-segment
with respect to the shower-axis versus its total energy.  Although
high energy particles are very much aligned with the shower axis,
there is an isotropic low energy component which moves with the
shower.  The isotropic component of the shower does not contribute to
the Cherenkov pulse, which validates the removal of the particles when
they fall below a certain low-energy threshold, say 0.611 MeV.  }
\end{figure}

Our study shows (see Fig. \ref{fig:polar-angle}) that there is indeed
an isotropic component in the shower below energy 0.85 MeV.  However,
they do not contribute to the radiation at the Cherenkov angle
($\theta_{\rm c} \approx 55.8^{\circ}$).  Particles with energy 1 MeV,
for example, Cherenkov radiates at an angle $49^{\circ}$.

\subsection{\it Angular Pulse Distribution}

\begin{figure}
\vskip 6.75cm
\center
\begin{picture}(0,0)
\includegraphics{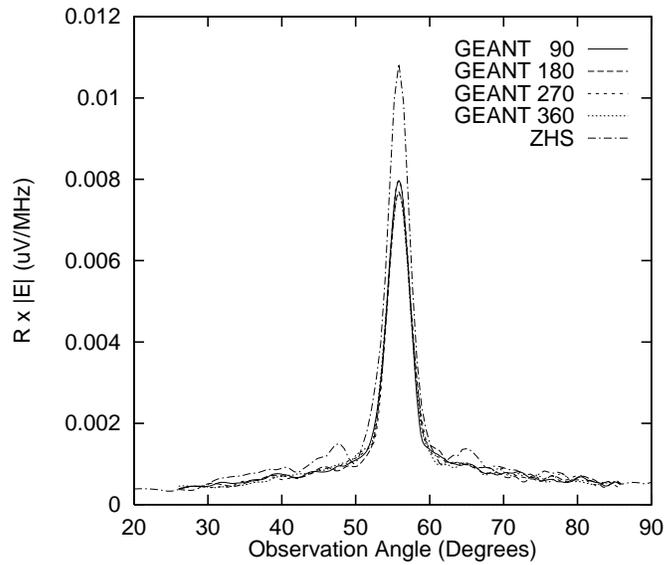}   
\end{picture}
\vskip 0.5cm
\caption{\label{fig:ps1ghz}
         \small
Angular pulse distribution of a 100 GeV shower (averaged over 50
showers) with 0.611 MeV total energy threshold from GEANT.  The pulse
is calculated at 1 GHz frequency and at 4 different azimuthal angles
($\phi$). The Cherenkov peak at the observation angle $\theta =
\theta_{\rm c}$ shows very little dependence on $\phi$.  The same
pulse from the ZHS Monte Carlo is also plotted for comparison.}
\end{figure}

The angular distribution of the electric field amplitude peaks at the
Cherenkov angle ($\theta_{\rm c} \approx 55.8^{\circ}$).
Fig. \ref{fig:ps1ghz} shows the 1 GHz pulse from a 100 GeV shower
(averaged over 50 showers) with 0.611 MeV total energy threshold at 4
different azimuthal angles $\phi = 90^{\circ},\; 180^{\circ},\;
270^{\circ}\; {\rm and}\; 360^{\circ}$.  The pulse shows very little
dependence on $\phi$ which corresponds to the approximately symmetric
distribution of particles about the shower axis as stated earlier.  We
have also plotted the electric field amplitude from the ZHS Monte
Carlo for a 100 GeV shower (averaged over 50 showers) at 1 GHz
frequency in Fig. \ref{fig:ps1ghz} for comparison.

\begin{figure}
\vskip 6.75cm
\center
\begin{picture}(0,0)
\includegraphics{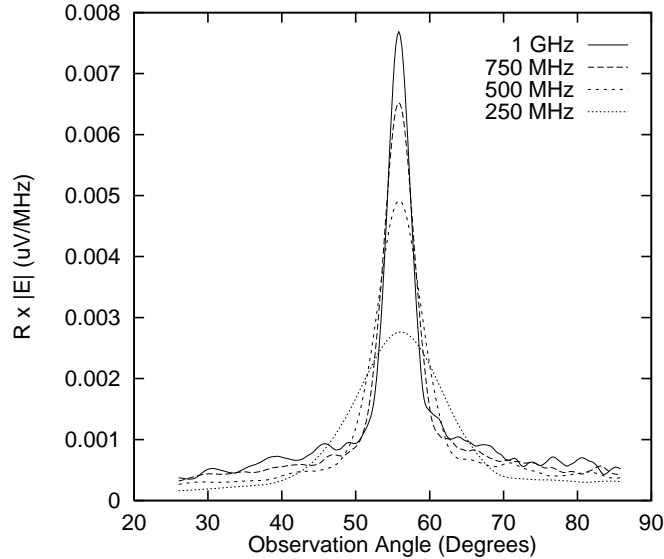}   
\end{picture}
\vskip 0.5cm
\caption{\label{fig:psmulti}
         \small
Angular pulse distribution of a 100 GeV shower (averaged over 50
showers) with 0.611 MeV total energy threshold from GEANT.  The pulse
is calculated at frequencies: 1 GHz, 750 MHz, 500 MHz and 250 MHz.
The plot shows inverse scaling of Gaussian half width of the Cherenkov
pulse with frequency.}
\end{figure}

Fig. \ref{fig:psmulti} shows the angular pulse distribution for a 100
GeV shower (averaged over 50 showers) at 1 GHz, 750 MHz, 500 MHz and
250 MHz frequencies. The Gaussian half width of the pulses are
2.0$^{\circ}$, 3.7$^{\circ}$ and 7.3$^{\circ}$ at 1 GHz, 500 MHz and
250 MHz frequencies respectively. This corresponds to approximate
inverse scaling of pulse width with frequency which is analogous to a
single-slit diffraction pattern as pointed out in reference
\cite{zhs92}.

\section{Summary of Results and Conclusions}

We have analyzed 100 GeV - 1 TeV electromagnetic showers and the radio
frequency radiation they produce in great detail.  These studies are a
necessary ingredient for designing an experiment for radio detection
of UHE cosmic ray induced showers in radio- transparent media.  In
particular, experiments to detect radio emission from showers induced
by high energy cosmic ray neutrinos interacting in the surface of the
moon \cite{gorham01} and in the South Polar ice-cap \cite{rice01} are
underway and have reported preliminary results.  Coherent radio
emission from electromagnetic showers has recently been demonstrated
in the laboratory \cite{saltzberg01}.  The technique is gaining
recognition as a powerful tool for particle detection. The thorough
dissection and understanding of all the intricacies of the showers and
their relationships to the final radio pulse produced is our goal
achieved in this paper.

Energy information for each stage of every track is readily available
from the GEANT shower code.  This information is essential for
tracking the energy distribution in the shower and identifying the
sources of radio emission.  Among our results, is the {\it direct}
determination of the radiation length in ice from an exponential fit
to bremsstrahlung radiation energy loss as a function of depth in
Sec. 3.1.  When this information is combined with the {\it direct}
extraction of the ionization loss, Sec. 3.3, we determined the {\it
critical energy} from the data shown in Fig. \ref{fig:critenergy}.
The value obtained is nicely consistent with that found by using the
{\it Moliere} radius extracted from the data for the radial energy
flow, Sec. 3.2, in combination with the radiation length.  The
consistency between the critical energy values indicates that our
application of the code and analysis of the data gives a correct
physical picture of the interplay among the competing processes in the
shower as it develops.

Because the coherent radio emission of interest depends upon the
charge excess in the shower, we need the energy profile of the
contributions to the charge excess. This again requires the GEANT
track-by-track energy information, and we show the total charge
imbalance broken down into energy ranges in Fig. \ref{fig:depfrac}.  A
large fraction, more than 50\%, of the imbalance comes from the energy
range below 5 MeV.  Though the track lengths are small, the large
number of particles leads to a significant contribution to the total
track length of the shower.

The total and projected track lengths for both the total and excess
charge populations are determined and shown to be proportional to
total shower energy in Sec. 3.5.  The total track length is
comfortably below a rough estimate of the upper bound for the total
track length.  Our detailed study of the longitudinal profiles in iron
shows overall agreement but differences in detail among the GEANT, the
EGS4 and the ZHS simulations in Fig. \ref{fig:depiron}.  The profiles
in ice for 100 GeV, 500 GeV and 1 TeV electron and photon induced
showers in ice are well described by a modified Greisen
parametrization with critical energy value extracted from the
simulation data, as described above, and two fit parameters.  The
confidence levels of the fits are typically 80\% - 90\%, as summarized
in Tables 1 \& 2.  The GEANT profiles in ice are qualitatively similar
to those from the ZHS code, but lie typically 25\% - 35\% lower.  To
determine whether small differences in cross section values used in
the different simulations could account for this difference, we
developed a 1-dimensional shower code with the full set of cross
sections for the relevant processes.  As we report in Sec. 3.7, the
depth at maximum and the number of particles at maximum are rather
insensitive to changes in the cross sections. We therefore believe
that the differences in cross-sections are not the source of profile
and total track length differences between our simulation and that of
ZHS.  Unfortunately, one important check that we have not been able to
make is the GEANT vs. ZHS ionization energy loss ($\dd E/\dd x$). The
ZHS code does not admit a readout of energy loss by shower
particles. Therefore, we were unable to make from ZHS code plots
similar to Figs. \ref{fig:dedxdat} \& \ref{fig:dedxavg} we made with
GEANT.  If the ionization loss in ZHS code is much lower than in
GEANT, it might account for the difference between showers produced by
the two Monte Carlos.

We develop the framework to calculate the electric field in the 100
MHz to multi GHz frequency range in Sec. 5.  The track-by-track field
calculation, applicable in the Fraunhoffer zone, and the calculation
treating the shower as a continuous current density, applicable in the
Fraunhoffer {\it and} and Fresnel zones, complement each other.  We
present both for this reason.  The direct, numerical calculation of
the total field from the vector sum of the fields from individual
tracks allows a detailed study of the dependence on total track
length.  Moreover, we elucidate the effects of random direction
changes due to collisions and the pattern of phase relationships among
the contributions from all tracks.  Complementing these intensive
numerical calculations, we present analytic work that employs the {\it
form factor} of the effective current density to calculate the field
in both Fresnel and Fraunhoffer regimes.  Given our realistic model
for the current density, which we fit to the transverse charge
distribution from the simulation, we show the remarkable result that
the radiation from the shower is {\it coherent over the whole shower}
at the Cherenkov angle in the Fraunhoffer limit.

Our study of the track-by-track phase coherence in Sec. 6 reveals that
the phase associated with the initial time and position of a track
(TP) and the phase associated with the ``diffractive emission'' (CP)
from the track as a whole are uncorrelated.  This result supports our
model of the current, and consequently of the field, in Sec. 5.
Figs. \ref{fig:phase-c} \& \ref{fig:phase-40} show the strong phase
peaking of the ``diffractive'' phase, called Cherenkov phase (CP) in
the discussion, on and off the Cherenkov angle as defined by the
shower axis.  The phase associated with the initial coordinates of the
track, called the translation phase (TP), is coherent at the Cherenkov
angle but completely random away from that angle.

We noticed some time ago \cite{soebtalk00} that the frequency spectrum
of the electric field at the Cherenkov angle, calculated directly from
the track data and the Fraunhoffer zone formulas developed in Sec. 5.,
flattens out at frequencies above 2 GHz, as shown in Fig. 24.  It is
clear from the figure that the ZHS simulation shows the same behavior
but they did not examine the high frequency region further
\cite{zhs92}.  We employ the form factor, which correctly accounts for
the transverse spread of the shower and its effect on the field to
analyze the frequency spectrum.  We evaluate its Fourier transform,
$F(\omega)$, two independent ways and find agreement.  We then show
that the observed spectrum of the field is faithfully represented up
to 5 GHz for 100 GeV showers and up to 15 GHz for 500 GeV and 1 TeV
showers.  Up to 15 GHz, we now have a clear picture of the behavior of
the frequency spectrum, which continues to show coherent behavior.

The electric field in the Fraunhoffer zone as a function of the
observation angle from the shower axis peaks at the Cherenkov angle.
The width of the peak shrinks inversely with frequency and the height
(field strength) rises linearly with shower energy.  These features
confirm the ZHS results and the off-Cherenkov angular dependence is
also similar.  The height of the peak at the Cherenkov angle for a
given energy and frequency is larger in their case by about 25-35\%.
This is perhaps not a surprising difference between two independent
shower simulations and field calculations.  We have performed an
extensive set of tests and cross-checks to validate our results.

\subsection{Conclusions}

Our study quantifies coherent Cherenkov radiation from high energy
showers and shows that coherence persists from 100 MHz to tens of GHz.
The existence of Coherent Cherenkov radiation goes back to Askaryan,
and has been studied for decades, while the persistence of coherent
emission at the Cherenkov angle at multi-GHz frequencies is new.  The
persistence of coherence is established by our new, highly detailed
study of the actual shower currents and corresponding phase
distributions in CP and TP introduced in Sec. 6.  The recognition that
the TP is coherent over the whole length of the shower at the
Cherenkov angle is a new insight into the connection between shower
particles and the fields they produce.  We have analyzed the
electromagnetic shower characteristics in great detail, including the
energy distributions and energy losses, so that a complete set of
shower parameters can be extracted from the simulation data to confirm
and cross check the consistency of our picture.  A grand summary of
pertinent parameters and comparisons with the ZHS and other sources
where available is presented in Table 5.

\begin{center}
\begin{tabular}{lcc}
\multicolumn{3}{l}{Table 5: Comparisons between 100 GeV showers in
ice from GEANT} \\
\multicolumn{3}{l}{and from ZHS Monte Carlos}\\
\hline \hline
Parameter & GEANT &  ZHS \\ \hline
Total Energy Threshold (MeV) & 0.611 & 0.611 \\
Total absolute track length (meter) & 399 $\pm$ 5 & 642 \\
Total projected ($e+p$) track length (meter) & 374 $\pm$ 4
& 519 \\
Total projected ($e-p$) track length (meter) & 70 $\pm$ 8
& 131 \\   
Position of the shower max. (radiation length) & 6.5 & 7.0 \\
Number of particles ($e+p$) at shower max. & 111 $\pm$ 7
& 155 $\pm$ 10 \\
Excess electrons ($e-p$) at shower max. & 20 $\pm$ 2
& 37 $\pm$ 3 \\
Fractional charge excess at the shower max & $\sim$ 18\% & $\sim$ 24\%
\\
Cherenkov peak at 1 GHz (Volts/MHz) & 7.5 $\times$ 10$^{-9}$
& 1.0 $\times$ 10$^{-8}$ \\
\hline \hline
\end{tabular}
\end{center}

The simulation and field calculation developed are directly applicable
to energies up to 1 TeV.  Higher energy showers can be "bootstrapped"
by evolving the shower particles into this regime with a corresponding
multiplication in particle number and emitted field intensity.  We
intend to parametrize and extrapolate the data to the multi-PeV data
needed for the UHE showers relevant to the RICE experiment.  With the
GEANT base, it will also be interesting to investigate similar shower
and field calculation for the hadronic component of neutrino-induced
showers.

\subsection*{Acknowledgement}

Help and advice from Enrique Zas at various stages of this work were
invaluable.  Discussions and working sessions with Jaime Alvarez-Muniz
were essential to our understanding of the workings of the ZHS
code. We thank George Frichter and Florian Hardt for early diagnostic
works using the ZHS code.  Comments and questions from Frances Halzen
on this topic over many years have been most useful.  Thanks to Tim
Bolton for helping with initial set up of the GEANT Monte Carlo
code. This work is supported in part by the NSF, the DOE, the
University of Kansas General Research Fund, the RESEARCH CORPORATION
and the facilities of the Kansas Institute for Theoretical and
Computational Science.

\newcommand{\noopsort}[1]{} \newcommand{\printfirst}[2]{#1}
\newcommand{\singleletter}[1]{#1} \newcommand{\switchargs}[2]{#2#1}

\end{document}